\pdfoutput=1

\documentclass[11pt]{article}

\usepackage{acl}

\usepackage{times}
\usepackage{latexsym}

\usepackage[T1]{fontenc}

\usepackage[utf8]{inputenc}

\usepackage{microtype}
\usepackage{multirow}
\usepackage{booktabs}
\usepackage{xspace}
\usepackage{graphicx}
\usepackage{algorithm}
\usepackage{algpseudocode}
\usepackage{amsmath}
\usepackage{eqparbox}
\usepackage{todonotes}
\usepackage{subcaption}
\usepackage{siunitx}
\usepackage{accents}
\usepackage{inconsolata}
\usepackage{color,soul}

\newdimen{\algindent}
\algnewcommand\LeftComment[2]{%
\hspace{#1\algindent}$\triangleright$ \eqparbox{COMMENT}{#2} \hfill %
}
\newcommand{\secref}[1]{\S\ref{#1}\xspace}
\newcommand{\system}{ColBERTv2\xspace}
\newcommand{\dataset}{LoTTE\xspace}

\title{\system{}:
\\
Effective and Efficient Retrieval via Lightweight Late Interaction}

\author{Keshav Santhanam\Thanks{ Equal contribution.} \\ Stanford University \And Omar Khattab\footnotemark[1] \\ Stanford University \And Jon Saad-Falcon \\ Georgia Institute of Technology \AND Christopher Potts \\ Stanford University \And Matei Zaharia \\ Stanford University}

\begin{document}
\maketitle
\begin{abstract}
Neural information retrieval (IR) has greatly advanced search and other knowledge-intensive language tasks. While many neural IR methods encode queries and documents into single-vector representations, late interaction models produce multi-vector representations at the granularity of each token and decompose relevance modeling into scalable token-level computations. This decomposition has been shown to make late interaction more effective, but it inflates the space footprint of these models by an order of magnitude. In this work, we introduce \system{}, a retriever that couples an aggressive residual compression mechanism with a denoised supervision strategy to simultaneously improve the quality and space footprint of late interaction. We evaluate \system{} across a wide range of benchmarks, establishing state-of-the-art quality within and outside the training domain while reducing the space footprint of late interaction models by 6--10$\times$.

\end{abstract}

\section{Introduction} \label{sec:introduction}

Neural information retrieval (IR) has quickly dominated the search landscape over the past 2--3 years, dramatically advancing not only passage and document search \cite{nogueira2019passage} but also many knowledge-intensive NLP tasks like open-domain question answering \cite{guu2020realm}, multi-hop claim verification \cite{khattab2021baleen}, and open-ended generation \cite{paranjape2021hindsight}. 

Many neural IR methods follow a \textit{single-vector similarity} paradigm: a pretrained language model is used to encode each query and each document into a single high-dimensional vector, and relevance is modeled as a simple dot product between both vectors. An alternative is \textit{late interaction}, introduced in ColBERT~\cite{khattab2020colbert}, where queries and documents are encoded at a finer-granularity into multi-vector representations, and relevance is estimated using rich yet scalable interactions between these two sets of vectors. ColBERT produces an embedding for every token in the query (and document) and models relevance as the sum of maximum similarities between each query vector and all vectors in the document.

By decomposing relevance modeling into token-level computations, late interaction aims to reduce the burden on the encoder: whereas single-vector models must capture complex query--document relationships within one dot product, late interaction encodes meaning at the level of tokens and delegates query--document matching to the interaction mechanism. This added expressivity comes at a cost: existing late interaction systems impose an order-of-magnitude larger \textit{space footprint} than single-vector models, as they must store billions of small vectors for Web-scale collections. Considering this challenge, it might seem more fruitful to focus instead on addressing the fragility of single-vector models~\cite{menon2022in} by introducing new supervision paradigms for negative mining~\cite{xiong2020approximate}, pretraining~\cite{gao2021unsupervised}, and distillation~\cite{qu2021rocketqa}. Indeed, recent single-vector models with highly-tuned supervision strategies~\cite{ren2021rocketqav2,formal2021spladev2} sometimes perform on-par or even better than ``vanilla'' late interaction models, and it is not necessarily clear whether late interaction architectures---with their fixed token-level inductive biases---admit similarly large gains from improved supervision.

In this work, we show that late interaction retrievers naturally produce lightweight token representations that are amenable to efficient storage off-the-shelf and that they can benefit drastically from denoised supervision. 
We couple those in \textbf{\system{}},\footnote{Code, models, and \dataset{} data are maintained at \url{https://github.com/stanford-futuredata/ColBERT}}
a new late-interaction retriever that employs a simple combination of distillation from a cross-encoder and hard-negative mining (\secref{sec:supervision}) to boost quality beyond any existing method, and then uses a \textit{residual compression} mechanism (\secref{sec:representation}) to reduce the space footprint of late interaction by 6--10$\times$ while preserving quality. As a result, \system{} establishes state-of-the-art retrieval quality both \textit{within} and \textit{outside} its training domain with a competitive space footprint with typical single-vector models.

When trained on MS MARCO Passage Ranking, \system{} achieves the highest MRR@10 of any standalone retriever. In addition to in-domain quality, we seek a retriever that generalizes ``zero-shot'' to domain-specific corpora and long-tail topics, ones that are often under-represented in large public training sets. To this end, we evaluate \system{} on a wide array of \textit{out-of-domain} benchmarks. These include three Wikipedia Open-QA retrieval tests and 13 diverse retrieval and semantic-similarity tasks from BEIR~\cite{thakur2021beir}. In addition, we introduce a new benchmark, dubbed \textbf{\dataset{}}, for \underline{Lo}ng-\underline{T}ail \underline{T}opic-stratified \underline{E}valuation for IR that features 12 domain-specific search tests, spanning StackExchange communities and using queries from GooAQ~\cite{khashabi2021gooaq}. \dataset{} focuses on relatively long-tail topics in its passages, unlike the Open-QA tests and many of the BEIR tasks, and evaluates models on their capacity to answer natural search queries with a practical intent, unlike many of BEIR's semantic-similarity tasks. On 22 of 28 out-of-domain tests, \system{} achieves the highest quality, outperforming the next best retriever by up to 8\% relative gain, while using its compressed representations.

This work makes the following contributions:

\begin{enumerate}
    
    \item We propose \system{}, a retriever that combines denoised supervision and residual compression, leveraging the token-level decomposition of late interaction to achieve high robustness with a reduced space footprint.  %
    
    \item We introduce \dataset, a new resource for out-of-domain evaluation of retrievers. \dataset{} focuses on natural information-seeking queries over long-tail topics, an important yet understudied application space.
    
    \item We evaluate \system{} across a wide range of settings, establishing state-of-the-art quality within and outside the training domain.
\end{enumerate}

\section{Background \& Related Work} \label{sec:related_work}

\subsection{Token-Decomposed Scoring in Neural IR}

Many neural IR approaches encode passages as a single high-dimensional vector, trading off the higher quality of cross-encoders for improved efficiency and scalability~\cite{karpukhin2020dense, xiong2020approximate, qu2021rocketqa}. ColBERT's~\cite{khattab2020colbert} late interaction paradigm addresses this tradeoff by computing multi-vector embeddings and using a scalable ``MaxSim'' operator for retrieval. Several other systems leverage multi-vector representations, including Poly-encoders~\cite{humeau2020polyencoders}, PreTTR~\cite{macavaney2020efficient}, and MORES~\cite{gao2020modularized}, but these target attention-based re-ranking as opposed to ColBERT's scalable MaxSim end-to-end retrieval. %

ME-BERT~\cite{luan2021sparse} generates token-level document embeddings similar to ColBERT, but retains a single embedding vector for queries. COIL~\cite{gao2021coil} also generates token-level document embeddings, but the token interactions are restricted to lexical matching between query and document terms. uniCOIL~\cite{lin2021few} limits the token embedding vectors of COIL to a single dimension, reducing them to scalar weights that extend models like DeepCT~\cite{dai2020context} and DeepImpact~\cite{mallia2021learning}. To produce scalar weights, SPLADE~\cite{formal2021splade} and SPLADEv2~\cite{formal2021spladev2} produce a sparse vocabulary-level vector that retains the term-level decomposition of late interaction while simplifying the storage into one dimension per token. The SPLADE family also piggybacks on the language modeling capacity acquired by BERT during pretraining. SPLADEv2 has been shown to be highly effective, within and across domains, and it is a central point of comparison in the experiments we report on in this paper.

\subsection{Vector Compression for Neural IR}

There has been a surge of recent interest in compressing representations for IR. \citet{izacard2020memory} explore dimension reduction, product quantization (PQ), and passage filtering for single-vector retrievers. BPR~\cite{yamada2021efficient} learns to directly hash embeddings to binary codes using a differentiable \texttt{tanh} function. JPQ~\cite{zhan2021jointly} and its extension, RepCONC~\cite{zhan2021learning}, use PQ to compress embeddings, and jointly train the query encoder along with the centroids produced by PQ via a ranking-oriented loss.

SDR~\cite{cohen2021sdr} uses an autoencoder to reduce the dimensionality of the contextual embeddings used for attention-based re-ranking and then applies a quantization scheme for further compression. DensePhrases~\cite{lee-etal-2021-learning-dense} is a system for Open-QA that relies on a multi-vector encoding of passages, though its search is conducted at the level of individual vectors and not aggregated with late interaction. Very recently, \citet{lee2021phrase} propose a quantization-aware finetuning method based on PQ to reduce the space footprint of DensePhrases. While DensePhrases is effective at Open-QA, its retrieval quality---as measured by top-20 retrieval accuracy on NaturalQuestions and TriviaQA---is competitive with DPR~\cite{karpukhin2020dense} and considerably less effective than ColBERT~\cite{khattab2021relevance}.

In this work, we focus on late-interaction retrieval and investigate compression using a residual compression approach that can be applied off-the-shelf to late interaction models, without special training. We show in Appendix~\ref{sec:analysis} that ColBERT's representations naturally lend themselves to residual compression. Techniques in the family of residual compression are well-studied~\cite{barnes1996advances} and have previously been applied across several domains, including approximate nearest neighbor search~\cite{wei2014projected, ai2017optimized}, neural network parameter and activation quantization~\cite{li2021residual, li2021trq}, and distributed deep learning~\cite{chen2018adacomp, liu2020double}. To the best of our knowledge, \system{} is the first approach to use residual compression for scalable neural IR.

\subsection{Improving the Quality of Single-Vector Representations}
\label{sec:related:supervision}

Instead of compressing multi-vector representations as we do, much recent work has focused on improving the quality of single-vector models, which are often very sensitive to the specifics of supervision. This line of work can be decomposed into three directions: (1) distillation of more expressive architectures~\cite{hofstatter2020improving, lin2020distilling} including explicit denoising~\cite{qu2021rocketqa,ren2021rocketqav2}, (2) hard negative sampling~\cite{xiong2020approximate,zhan2020learning,zhan2021optimizing}, and (3) improved pretraining~\cite{gao2021unsupervised, ouguz2021domain}. We adopt similar techniques to (1) and (2) for \system{}'s multi-vector representations (see \secref{sec:supervision}).

\subsection{Out-of-Domain Evaluation in IR}
\label{sec:related:ood}

Recent progress in retrieval has mostly focused on large-data evaluation, where many tens of thousands of annotated training queries are associated with the test domain, as in MS MARCO or Natural Questions \citep{kwiatkowski2019natural}. In these benchmarks, queries tend to reflect high-popularity topics like movies and athletes in Wikipedia. In practice, user-facing IR and QA applications often pertain to domain-specific corpora, for which little to no training data is available and whose topics are under-represented in large public collections.

This out-of-domain regime has received recent attention with the BEIR~\cite{thakur2021beir} benchmark. BEIR combines several existing datasets into a heterogeneous suite for ``zero-shot IR'' tasks, spanning bio-medical, financial, and scientific domains. While the BEIR datasets provide a useful testbed, many capture broad semantic relatedness tasks---like citations, counter arguments, or duplicate questions--instead of natural search tasks, or else they focus on high-popularity entities like those in Wikipedia. In \secref{section:sedata}, we introduce \dataset{}, a new dataset for out-of-domain retrieval, exhibiting natural search queries over long-tail topics.

\begin{figure}[]
\centering
\includegraphics[width=0.85\columnwidth]{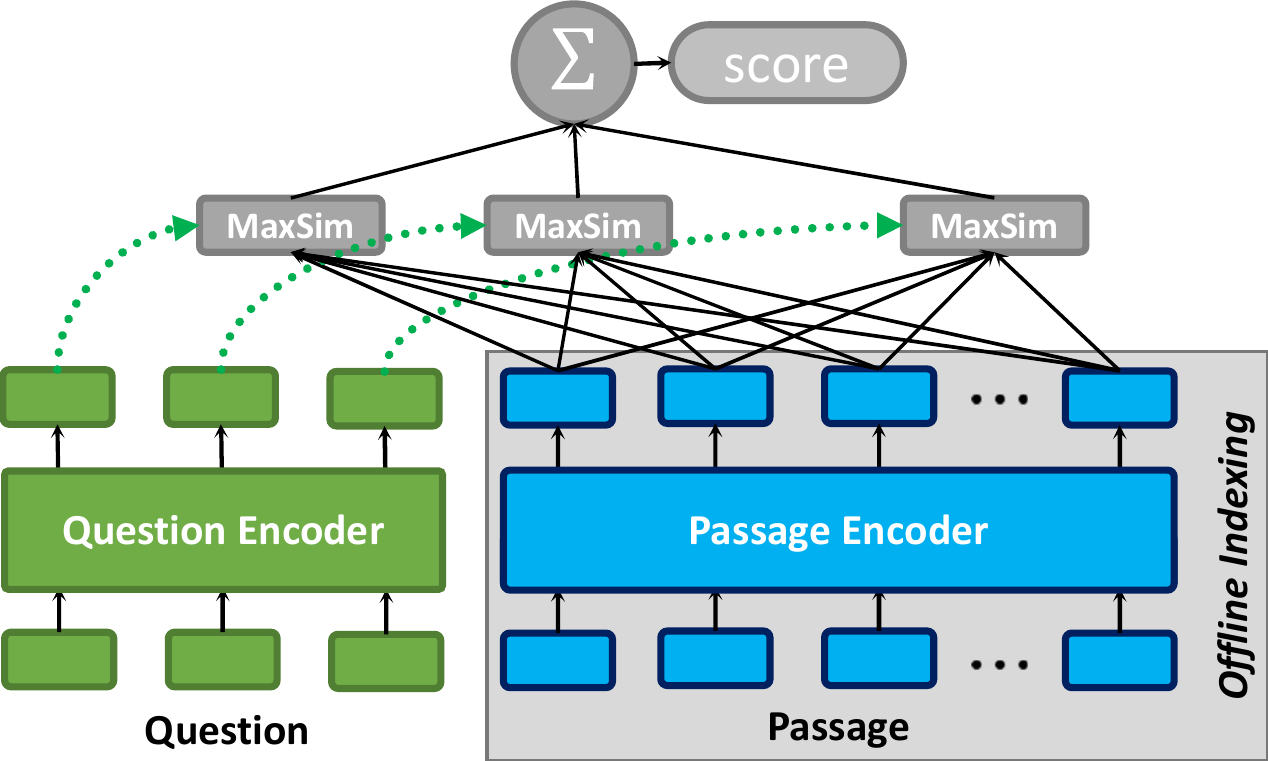}
\vspace*{-2mm}
\caption{The late interaction architecture, given a query and a passage. Diagram from \citet{khattab2021relevance} with permission.}
\label{fig:ColBERT}
\vspace{-3mm}
\end{figure}

\section{\system} \label{sec:maize}

We now introduce \system{}, which improves the quality of multi-vector retrieval models (\secref{sec:supervision}) while reducing their space footprint (\secref{sec:representation}).

\subsection{Modeling}  \label{sec:modeling}

\system{} adopts the late interaction architecture of ColBERT, depicted in Figure~\ref{fig:ColBERT}. Queries and passages are independently encoded with BERT~\cite{devlin-etal-2019-bert}, and the output embeddings encoding each token are projected to a lower dimension. During offline indexing, every passage $d$ in the corpus is encoded into a set of vectors, and these vectors are stored. At search time, the query $q$ is encoded into a multi-vector representation, and its similarity to a passage $d$ is computed as the summation of query-side ``MaxSim'' operations, namely, the largest cosine similarity between each query token embedding and all passage token embeddings:
\begin{equation}
\label{eq:scorer}
S_{q,d} = \sum_{i=1}^{N} \max_{j=1}^{M} Q_{i} \cdot D_{j}^{T}
\end{equation}
where $Q$ is an matrix encoding the query with $N$ vectors and $D$ encodes the passage with $M$ vectors. The intuition of this architecture is to align each query token with the most contextually relevant passage token, quantify these matches, and combine the partial scores across the query. We refer to \citet{khattab2020colbert} for a more detailed treatment of late interaction.

\subsection{Supervision} \label{sec:supervision}

Training a neural retriever typically requires \textit{positive} and \textit{negative} passages for each query in the training set. \citet{khattab2020colbert} train ColBERT using the official $\langle$q, d$^+$, d$^-\rangle$ triples of MS~MARCO. For each query, a positive $d^+$ is human-annotated, and each negative $d^-$ is sampled from unannotated BM25-retrieved passages.

Subsequent work has identified several weaknesses in this standard supervision approach (see~\secref{sec:related:supervision}). Our goal is to adopt a simple, uniform supervision scheme that selects challenging negatives and avoids rewarding false positives or penalizing false negatives. To this end, we start with a ColBERT model trained with triples as in \citet{khattab2021relevance}, using this to index the training passages with \system{} compression.

For each training query, we retrieve the top-$k$ passages. We feed each of those query--passage pairs into a cross-encoder reranker. We use a 22M-parameter MiniLM~\cite{wang2020minilm} cross-encoder trained with distillation by \citet{thakur2021beir}.\footnote{\url{https://huggingface.co/cross-encoder/ms-marco-MiniLM-L-6-v2}} This small model has been shown to exhibit very strong performance while being relatively efficient for inference, making it suitable for distillation.

We then collect $w$-way tuples consisting of a query, a highly-ranked passage (or labeled positive), and one or more lower-ranked passages. In this work, we use $w=64$ passages per example. Like RocketQAv2~\cite{ren2021rocketqav2}, we use a KL-Divergence loss to distill the cross-encoder's scores into the ColBERT architecture. We use KL-Divergence as ColBERT produces scores (i.e., the sum of cosine similarities) with a restricted scale, which may not align directly with the output scores of the cross-encoder. We also employ in-batch negatives per GPU, where a cross-entropy loss is applied to the positive score of each query against all passages corresponding to other queries in the same batch. We repeat this procedure once to refresh the index and thus the sampled negatives.

Denoised training with hard negatives has been positioned in recent work as ways to bridge the gap between single-vector and interaction-based models, including late interaction architectures like ColBERT. Our results in \secref{sec:evaluation} reveal that such supervision can improve multi-vector models dramatically, resulting in state-of-the-art retrieval quality.

\subsection{Representation} \label{sec:representation}

We hypothesize that the ColBERT vectors cluster into regions that capture highly-specific token semantics. We test this hypothesis in Appendix~\ref{sec:analysis}, where evidence suggests that vectors corresponding to each sense of a word cluster closely, with only minor variation due to context. We exploit this regularity with a \textit{residual} representation that dramatically reduces the space footprint of late interaction models, completely \textit{off-the-shelf} without architectural or training changes. Given a set of centroids $C$, \system{} encodes each vector $v$ as the index of its closest centroid $C_t$ and a \textit{quantized} vector $\tilde{r}$ that approximates the residual $r = v - C_t$. At search time, we use the centroid index $t$ and residual $\tilde{r}$ recover an approximate $\tilde{v} = C_t + \tilde{r}$.

To encode $\tilde{r}$, we quantize every dimension of $r$ into one or two bits. In principle, our $b$-bit encoding of $n$-dimensional vectors needs $\lceil \log{|C|} \rceil+b n$ bits per vector. In practice, with $n=128$, we use four bytes to capture up to $2^{32}$ centroids and 16 or 32 bytes (for $b=1$ or $b=2$) to encode the residual. This total of 20 or 36 bytes per vector contrasts with ColBERT's use of 256-byte vector encodings at 16-bit precision. While many alternatives can be explored for compression, we find that this simple encoding largely preserves model quality, while considerably lowering storage costs against typical 32- or 16-bit precision used by existing late interaction systems.

This centroid-based encoding can be considered a natural extension of product quantization to \textit{multi-vector} representations. Product quantization~\cite{gray1984vector,jegou2010product} compresses a single vector by splitting it into small sub-vectors and encoding each of them using an ID within a codebook. In our approach, each representation is already a matrix that is naturally divided into a number of small vectors (one per token). We encode each vector using its nearest centroid plus a residual. Refer to Appendix~\ref{appendix:compression} for tests of the impact of compression on retrieval quality and a comparison with a baseline compression method for ColBERT akin to BPR~\cite{yamada-etal-2021-efficient}.

\begin{table*}[]
\small
\resizebox{\linewidth}{!}{
\begin{tabular}{@{}lrrrcrrc@{}}
\toprule
\multicolumn{1}{c}{\multirow{2}{*}{Topic}} & \multicolumn{1}{c}{\multirow{2}{*}{Question Set}} & \multicolumn{3}{c}{Dev}                                                                                                                                         & \multicolumn{3}{c}{Test}                                                                                                                                       \\ \cmidrule(l){3-8} 
\multicolumn{1}{c}{}                       & \multicolumn{1}{c}{}                              & \multicolumn{1}{c}{\# Questions} & \multicolumn{1}{c}{\# Passages} & Subtopics                                                                                  & \multicolumn{1}{c}{\# Questions} & \multicolumn{1}{c}{\# Passages} & Subtopics                                                                                 \\ \toprule
\multirow{2}{*}{Writing}                   & Search                                            & 497                              & \multirow{2}{*}{277k}         & \multirow{2}{*}{\begin{tabular}[c]{@{}c@{}}ESL, Linguistics,\\ Worldbuilding\end{tabular}} & 1071                             & \multirow{2}{*}{200k}         & \multirow{2}{*}{English}                                                                  \\
                                           & Forum                                 & 2003                             &                                 &                                                                                            & 2000                             &                                 &                                                                                           \\  \midrule
\multirow{2}{*}{Recreation}             & Search                                            & 563                              & \multirow{2}{*}{263k}         & \multirow{2}{*}{\begin{tabular}[c]{@{}c@{}}Sci-Fi, RPGs,\\ Photography\end{tabular}}       & 924                              & \multirow{2}{*}{167k}         & \multirow{2}{*}{\begin{tabular}[c]{@{}c@{}}Gaming, \\Anime, Movies\end{tabular}}          \\
                                           & Forum                                 & 2002                             &                                 &                                                                                            & 2002                             &                                 &                                                                                           \\ \midrule
\multirow{2}{*}{Science}                 & Search                                            & 538                              & \multirow{2}{*}{344k}         & \multirow{2}{*}{\begin{tabular}[c]{@{}c@{}}Chemistry,\\ Statistics, Academia\end{tabular}} & 617                              & \multirow{2}{*}{1.694M}        & \multirow{2}{*}{\begin{tabular}[c]{@{}c@{}}Math, \\Physics, Biology\end{tabular}}         \\
                                           & Forum                                 & 2013                             &                                 &                                                                                            & 2017                             &                                 &                                                                                           \\ \midrule
\multirow{2}{*}{Technology}                & Search                                            & 916                              & \multirow{2}{*}{1.276M}        & \multirow{2}{*}{\begin{tabular}[c]{@{}c@{}}Web Apps,\\ Ubuntu, SysAdmin\end{tabular}}      & 596                              & \multirow{2}{*}{639k}         & \multirow{2}{*}{\begin{tabular}[c]{@{}c@{}}Apple, Android,\\ UNIX, Security\end{tabular}} \\
                                           & Forum                                 & 2003                             &                                 &                                                                                            & 2004                             &                                 &                                                                                           \\ \midrule
\multirow{2}{*}{Lifestyle}                 & Search                                            & 417                              & \multirow{2}{*}{269k}         & \multirow{2}{*}{\begin{tabular}[c]{@{}c@{}}DIY, Music, Bicycles,\\Car Maintenance\end{tabular}}            & 661                              & \multirow{2}{*}{119k}         & \multirow{2}{*}{\begin{tabular}[c]{@{}c@{}}Cooking, \\Sports, Travel\end{tabular}}        \\
                                           & Forum                                 & 2076                             &                                 &                                                                                            & 2002                             &                                 &                                                                                           \\ \midrule
\multirow{2}{*}{Pooled}                 & Search                                            & 2931                              & \multirow{2}{*}{2.4M}         & \multirow{2}{*}{\begin{tabular}[c]{@{}c@{}}All of the above\end{tabular}}            & 3869                              & \multirow{2}{*}{2.8M}         & \multirow{2}{*}{\begin{tabular}[c]{@{}c@{}}All of the above\end{tabular}}        \\
                                           & Forum                                 & 10097                             &                                 &                                                                                            & 10025                             &                                 &                                                                                           \\ 
                                           \bottomrule
\end{tabular}
}
\vspace{-2mm}
\caption{Composition of \dataset showing topics, question sets, and a sample of corresponding subtopics. Search Queries are taken from GooAQ, while Forum Queries are taken directly from the StackExchange archive. The pooled datasets combine the questions and passages from each of the subtopics.}
\label{table:stackexchange_composition}
\vspace{-3mm}
\end{table*}

\subsection{Indexing} \label{sec:indexing}

Given a corpus of passages, the indexing stage precomputes all passage embeddings and organizes their representations to support fast nearest-neighbor search. \system{} divides indexing into three stages, described below.

\textbf{Centroid Selection.} In the first stage, \system{} selects a set of cluster centroids $C$. These are embeddings that \system{} uses to support residual encoding (\secref{sec:representation}) and also for nearest-neighbor search (\secref{sec:retrieval}). Standardly, we find that setting $|C|$ proportionally to the square root of $n_{\text{embeddings}}$ in the corpus works well empirically.\footnote{We round down to the nearest power of two larger than $16 \times \sqrt{n_{\text{embeddings}}}$, inspired by FAISS~\cite{johnson2019billion}.} \citet{khattab2020colbert} only clustered the vectors after computing the representations of all passages, but doing so requires storing them uncompressed. To reduce memory consumption, we apply $k$-means clustering to the embeddings produced by invoking our BERT encoder over only a sample of all passages, proportional to the square root of the collection size, an approach we found to perform well in practice.

\textbf{Passage Encoding.} Having selected the centroids, we encode every passage in the corpus. This entails invoking the BERT encoder and compressing the output embeddings as described in \secref{sec:representation}, assigning each embedding to the nearest centroid and computing a quantized residual. Once a chunk of passages is encoded, the \textit{compressed} representations are saved to disk.

\textbf{Index Inversion.} To support fast nearest-neighbor search, we group the embedding IDs that correspond to each centroid together, and save this \textit{inverted list} to disk. At search time, this allows us to quickly find token-level embeddings similar to those in a query.

\subsection{Retrieval} \label{sec:retrieval}

Given a query representation $Q$, retrieval starts with candidate generation. For every vector $Q_i$ in the query, the nearest $n_\text{probe} \geq 1$ centroids are found. Using the inverted list, \system{} identifies the passage embeddings close to these centroids, decompresses them, and computes their cosine similarity with every query vector. The scores are then grouped by passage ID for each query vector, and scores corresponding to the same passage are $\max$-reduced. This allows \system{} to conduct an approximate ``MaxSim'' operation per query vector. This computes a lower-bound on the true MaxSim (\secref{sec:modeling}) using the embeddings identified via the inverted list, which resembles the approximation explored for scoring by \citet{macdonald2021approximate} but is applied for candidate generation.

These lower bounds are summed across the query tokens, and the top-scoring $n_\text{candidate}$ candidate passages based on these approximate scores are selected for ranking, which loads the complete set of embeddings of each passage, and conducts the same scoring function using all embeddings per document following Equation~\ref{eq:scorer}. The result passages are then sorted by score and returned.

\section{\dataset{}: Long-Tail, Cross-Domain Retrieval Evaluation} \label{section:sedata}

We introduce \textbf{\dataset{}} (pronounced latte), a new dataset for \textbf{Lo}ng-\textbf{T}ail \textbf{T}opic-stratified \textbf{E}valuation for IR. To complement the out-of-domain tests of BEIR~\cite{thakur2021beir}, as motivated in \secref{sec:related:ood}, \dataset{} focuses on \textit{natural user queries} that pertain to \textit{long-tail topics}, ones that might not be covered by an entity-centric knowledge base like Wikipedia. \dataset{} consists of 12 test sets, each with 500--2000 queries and 100k--2M passages.

The test sets are explicitly divided by topic, and each test set is accompanied by a validation set of \textit{related but disjoint} queries \textit{and} passages. We elect to make the passage texts disjoint to encourage more realistic out-of-domain transfer tests, allowing for minimal development on related but distinct topics. The test (and dev) sets include a ``pooled'' setting. In the pooled setting, the passages and queries are aggregated across all test (or dev) topics to evaluate out-of-domain retrieval across a larger and more diverse corpus.

Table~\ref{table:stackexchange_composition} outlines the composition of \dataset{}. We derive the topics and passage corpora from the \textit{answer posts} across various StackExchange forums. StackExchange is a set of question-and-answer communities that target individual topics (e.g., ``physics'' or ``bicycling''). We gather forums from five overarching domains: writing, recreation, science, technology, and lifestyle. To evaluate retrievers, we collect \textit{Search} and \textit{Forum} queries, each of which is associated with one or more target answer posts in its corpus. Example queries, and short snippets from posts that answer them in the corpora, are shown in Table~\ref{table:se-examples}. %

\textbf{Search Queries.} We collect search queries from GooAQ~\cite{khashabi2021gooaq}, a recent dataset of Google search-autocomplete queries and their answer boxes, which we filter for queries whose answers link to a specific StackExchange post. As \citet{khashabi2021gooaq} hypothesize, Google Search likely maps these natural queries to their answers by relying on a wide variety of signals for relevance, including expert annotations, user clicks, and hyperlinks as well as specialized QA components for various question types \textit{with access to the post title and question body}. Using those annotations as ground truth, we evaluate the models on their capacity for retrieval using \textit{only} free text of the answer posts (i.e., no hyperlinks or user clicks, question title or body, etc.), posing a significant challenge for IR and NLP systems trained only on public datasets.

\begin{table}[tp]

\centering
\small
\begin{tabular}{p{0.92\columnwidth}}
\toprule
  \textbf{Q:} \textit{what is the difference between root and stem in linguistics?}
  \textbf{A:} \textcolor{darkgray}{\textcolor{gray}{A root is }\textbf{the form to which derivational affixes are added } \textcolor{gray}{to form a stem. A stem is }\textbf{the form to which inflectional affixes are added} \textcolor{gray}{to form a word.}} \\ \midrule
  \textbf{Q:} \textit{are there any airbenders left?}
  \textbf{A:}  \textcolor{darkgray}{\textcolor{gray}{the Fire Nation had wiped out all Airbenders while Aang was frozen.} \textbf{Tenzin and his 3 children are the only Airbenders left in Korra's time.}}
 \\ \midrule
   \textbf{Q:} \textit{Why are there two Hydrogen atoms on some periodic tables?}
   \textbf{A:} \textcolor{darkgray}{\textcolor{gray}{some periodic tables show hydrogen in both places }\textbf{to emphasize that hydrogen isn't really a member of the first group or the seventh group.}}  \\
   \midrule
  \textbf{Q:} \textit{How can cache be that fast?}
  \textbf{A:} \textcolor{darkgray}{\textcolor{gray}{the cache memory sits right next to the CPU on the same die (chip), }\textbf{it is made using SRAM which is much, much faster than the DRAM.}} %
 \\ \bottomrule
\end{tabular}%

\vspace{0mm}
\caption{Examples of queries and shortened snippets of answer passages from \dataset{}. The first two examples show ``search'' queries, whereas the last two are ``forum'' queries. Snippets are shortened for presentation.}

\label{table:se-examples}
\vspace{-2mm}
\end{table}
\begin{table}[tp]
\small
\centering
\begin{tabular}{p{\columnwidth}}
\toprule
  \textbf{Q:} \textit{what is xerror in rpart?} \hspace{4mm}
  \textbf{Q:} \textit{is sub question one word?} \hspace{4mm}
  \textbf{Q:} \textit{how to open a garage door without making noise?} \hspace{4mm}
  \textbf{Q:} \textit{is docx and dotx the same?} \hspace{4mm}
  \textbf{Q:} \textit{are upvotes and downvotes anonymous?} \hspace{4mm}
  \textbf{Q:} \textit{what is the difference between descriptive essay and narrative essay?} \hspace{4mm}
  \textbf{Q:} \textit{how to change default user profile in chrome?} \hspace{4mm}
  \textbf{Q:} \textit{does autohotkey need to be installed?} \hspace{4mm}
  \textbf{Q:} \textit{how do you tag someone on facebook with a youtube video?} \hspace{4mm}
  \textbf{Q:} \textit{has mjolnir ever been broken?}
 \\ \midrule 
   \textbf{Q:} \textit{Snoopy can balance on an edge atop his doghouse. Is any reason given for this?} \hspace{4mm}
   \textbf{Q:} \textit{How many Ents were at the Entmoot?} \hspace{4mm}
  \textbf{Q:} \textit{What does a hexagonal sun tell us about the camera lens/sensor?} \hspace{4mm}
  \textbf{Q:} \textit{Should I simply ignore it if authors assume that Im male in their response to my review of their article?} \hspace{4mm}
  \textbf{Q:} \textit{Why is the 2s orbital lower in energy than the 2p orbital when the electrons in 2s are usually farther from the nucleus?} \hspace{4mm}
  \textbf{Q:} \textit{Are there reasons to use colour filters with digital cameras?} \hspace{4mm}
  \textbf{Q:} \textit{How does the current know how much to flow, before having seen the resistor?} \hspace{4mm}
  \textbf{Q:} \textit{What is the difference between Fact and Truth?} \hspace{4mm}
  \textbf{Q:} \textit{hAs a DM, how can I handle my Druid spying on everything with Wild shape as a spider?} \hspace{4mm}
  \textbf{Q:} \textit{What does 1x1 convolution mean in a neural network?}
 \\ \bottomrule
\end{tabular}%

\vspace{0mm}
\caption{Comparison of a random sample of search queries (top) vs. forum queries (bottom).}

\label{table:search_vs_forum_queries}
\end{table}

\textbf{Forum Queries.} We collect the forum queries by extracting post titles from the StackExchange communities to use as queries and collect their corresponding answer posts as targets. We select questions in order of their popularity and sample questions according to the proportional contribution of individual communities within each topic. These queries tend to have a wider variety than the ``search'' queries, while the search queries may exhibit more natural patterns. Table~\ref{table:search_vs_forum_queries} compares a random samples of search and forum queries. It can be seen that search queries tend to be brief, knowledge-based questions with direct answers, whereas forum queries tend to reflect more open-ended questions. Both query sets target topics that exceed the scope of a general-purpose knowledge repository such as Wikipedia.

For search as well as forum queries, the resulting evaluation set consists of a query and a target set of StackExchange answer posts (in particular, the answer posts from the target StackExchange page). Similar to evaluation in the Open-QA literature \cite{karpukhin2020dense,khattab2021relevance}, we evaluate retrieval quality by computing the \texttt{success@5} (S@5) metric. Specifically, we award a point to the system for each query where it finds an accepted or upvoted (score $\ge$ 1) answer from the target page in the top-5 hits. %

Appendix~\ref{section:appendix_lotte} reports on the breakdown of constituent communities per topic, the construction procedure of \dataset{} as well as licensing considerations, and relevant statistics. Figures~\ref{fig:lotte_words_per_query} and~\ref{fig:lotte_answers_per_query} quantitatively compare the search and forum queries.

\section{Evaluation} \label{sec:evaluation}

We now evaluate \system{} on passage retrieval tasks, testing its quality within the training domain (\secref{sec:in_domain}) as well as outside the training domain in zero-shot settings (\secref{sec:out_of_domain}).
Unless otherwise stated, we compress \system{} embeddings to $b=2$ bits per dimension in our evaluation. %

\begin{table}[tp]
\centering
\setlength{\tabcolsep}{3pt}
\resizebox{\columnwidth}{!}{
\begin{tabular}{@{}lrrrrrr@{}}
\toprule
\multicolumn{1}{c}{\multirow{2}{*}{Method}} & \multicolumn{3}{c}{Official Dev (7k)} & \multicolumn{3}{c}{Local Eval (5k)} \\
& \multicolumn{1}{c}{MRR@10} & \multicolumn{1}{c}{R@50}  & \multicolumn{1}{c}{R@1k}  &  \multicolumn{1}{c}{MRR@10} & \multicolumn{1}{c}{R@50}  & \multicolumn{1}{c}{R@1k}  \\ \midrule
              \multicolumn{7}{c}{Models without Distillation or Special Pretraining}               \\ \midrule
RepBERT             &    30.4 & - & 94.3 & - & - & -       \\
DPR     &       31.1 & - & 95.2 & - & - & -      \\
ANCE             &    33.0 & - & 95.9  & - & - & -    \\
LTRe              &   34.1   &   -   &  96.2 & - & - & -         \\
ColBERT             &     36.0 & 82.9 & 96.8 & 36.7 & - & -     \\
\midrule
              \multicolumn{7}{c}{Models with Distillation or Special Pretraining}               \\ \midrule
TAS-B                &   34.7    &    -      &       97.8  & - & - & -     \\
SPLADEv2            &   36.8    &   -       &       97.9 & 37.9 & 84.9 & 98.0   \\
PAIR                &   37.9    &   86.4    &       98.2 & - & - & -       \\
coCondenser          &   38.2    &    -      &       \textbf{98.4} & - & - & -      \\
RocketQAv2           &   38.8    &   86.2    &       98.1 & 39.8 & 85.8 & 97.9      \\    
\system{}             &   \textbf{39.7}    &   \textbf{86.8}    &       \textbf{98.4} & \textbf{40.8} & \textbf{86.3} & \textbf{98.3}      \\  \bottomrule

\end{tabular}
}
\vspace{-2mm}
\caption{In-domain performance on the development set of MS MARCO Passage Ranking as well the ``Local Eval'' test set described by \citet{khattab2020colbert}. Dev-set results for baseline systems are from their respective papers: \citet{zhan2020repbert}, \citet{xiong2020approximate} for DPR and ANCE, \citet{zhan2020learning}, \citet{khattab2020colbert}, \citet{hofstatter2021efficiently}, \citet{gao2021unsupervised}, \citet{ren2021pair}, \citet{formal2021spladev2}, and \citet{ren2021rocketqav2}.
}
\label{table:in_domain_results}
\vspace{-4mm}
\end{table}

\begin{table*}[]
\hspace{0.1cm}
\begin{subtable}[t]{0.45\linewidth}
\centering
\setlength{\tabcolsep}{4pt}
\small
\begin{tabular}[b]{@{}lcccc cccc@{}}
\toprule
\multicolumn{1}{c}{\multirow{2}{*}{Corpus}} & \multicolumn{4}{c}{Models without Distillation} & \multicolumn{4}{c}{Models with Distillation} \\ \cmidrule(l){2-9} 
                        & \rotatebox[origin=c]{270}{\scalebox{.85}{\textbf{ColBERT}}}      & \rotatebox[origin=c]{270}{\scalebox{.85}{\textbf{DPR-M}}}     & \rotatebox[origin=c]{270}{\scalebox{.85}{\textbf{ANCE}}}     & \rotatebox[origin=c]{270}{\scalebox{.85}{\textbf{MoDIR}}}     & \rotatebox[origin=c]{270}{\scalebox{.85}{\textbf{TAS-B}}}        &
                        \rotatebox[origin=c]{270}{\scalebox{.85}{\textbf{RocketQAv2}}}        &
                        \rotatebox[origin=c]{270}{\scalebox{.85}{\textbf{SPLADEv2}}}        & \rotatebox[origin=c]{270}{\scalebox{.85}{\textbf{\system{}}}}       \\ \midrule 
                           \multicolumn{8}{c}{BEIR Search Tasks (nDCG@10)}                                                                         \\ \midrule
\scalebox{.85}{\textbf{DBPedia}}                    & 39.2       & 23.6 & 28.1 & 28.4                & 38.4                      & 35.6     & 43.5                         & \textbf{44.6}             \\
\scalebox{.85}{\textbf{FiQA}}                  & 31.7       & 27.5 & 29.5 & 29.6                 & 30.0                      & 30.2     & 33.6                         & \textbf{35.6}             \\
\scalebox{.85}{\textbf{NQ}}                         & 52.4       & 39.8 & 44.6 & 44.2                 & 46.3                      & 50.5    & 52.1                         & \textbf{56.2}             \\
\scalebox{.85}{\textbf{HotpotQA}}                   & 59.3       & 37.1 & 45.6 & 46.2                 & 58.4                      & 53.3    & \textbf{68.4}                & 66.7                      \\
\scalebox{.85}{\textbf{NFCorpus}}                   & 30.5       & 20.8 & 23.7 & 24.4                 & 31.9                      & 29.3    & 33.4                         & \textbf{33.8}             \\
\scalebox{.85}{\textbf{T-COVID}}                 & 67.7       & 56.1 & 65.4 & 67.6                 & 48.1                      & 67.5   & 71.0                         & \textbf{73.8}             \\
\scalebox{.85}{\textbf{Touch\'e (v2)}}               & -      & - & - & -         & -                      & 24.7   & \textbf{27.2}                         & 26.3                      \\ \midrule
                           \multicolumn{8}{c}{BEIR Semantic Relatedness Tasks (nDCG@10)}                                                                 \\ \midrule
\scalebox{.85}{\textbf{ArguAna}}                    & 23.3          & 41.4 & 41.5 & 41.8              & 42.7                      & 45.1    & \textbf{47.9}                & 46.3                      \\
\scalebox{.85}{\textbf{C-FEVER}}              & 18.4       & 17.6 & 19.8 & 20.6                 & 22.8                      & 18.0      & \textbf{23.5}                & 17.6                      \\
\scalebox{.85}{\textbf{FEVER}}                      & 77.1       & 58.9 & 66.9 & 68.0                 & 70.0                      & 67.6    & \textbf{78.6}                & \textbf{78.5}             \\
\scalebox{.85}{\textbf{Quora}}                      & 85.4       & 84.2 & 85.2 & \textbf{85.6}            & 83.5                      & 74.9    & 83.8                         & 85.2                      \\
\scalebox{.85}{\textbf{SCIDOCS}}                    & 14.5       & 10.8 & 12.2 & 12.4                 & 14.9                      & 13.1    & \textbf{15.8}                & 15.4                      \\
\scalebox{.85}{\textbf{SciFact}}                    & 67.1       & 47.8 & 50.7 & 50.2                 & 64.3                      & 56.8    & \textbf{69.3}                & \textbf{69.3}             \\ \bottomrule
\end{tabular}
\caption{}
\label{table:beir_results}
\end{subtable} \hspace{1.8cm}
\begin{subtable}[t]{0.38\linewidth}
\centering
\footnotesize
\setlength{\tabcolsep}{4pt}
\begin{tabular}[b]{@{}lcccccc@{}}
\toprule
\multicolumn{1}{c}{Corpus} & \multicolumn{1}{c}{\rotatebox[origin=c]{270}{\scalebox{.85}{\textbf{ColBERT}}}} &
\multicolumn{1}{c}{\rotatebox[origin=c]{270}{\scalebox{.85}{\textbf{BM25}}}} &
\multicolumn{1}{c}{\rotatebox[origin=c]{270}{\scalebox{.85}{\textbf{ANCE}}}} &
\multicolumn{1}{c}{\rotatebox[origin=c]{270}{\scalebox{.85}{\textbf{RocketQAv2}}}} &
\multicolumn{1}{c}{\rotatebox[origin=c]{270}{\scalebox{.85}{\textbf{SPLADEv2}}}} & \multicolumn{1}{c}{\rotatebox[origin=c]{270}{\scalebox{.85}{\textbf{\system{}}}}} \\ \midrule 
                           \multicolumn{6}{c}{OOD Wikipedia Open QA (Success@5)}                                            \\ \midrule
\scalebox{.85}{\textbf{NQ}-dev}                         & 65.7                        & 44.6                         & -                      & -                      & 65.6                         & \textbf{68.9}             \\
\scalebox{.85}{\textbf{TQ}-dev}                         & 72.6                        & 67.6                         & -                      & -                      & 74.7                         & \textbf{76.7}             \\
\scalebox{.85}{\textbf{SQuAD}-dev}                      & 60.0                        & 50.6                         & -                      & -                      & 60.4                         & \textbf{65.0}             \\ \midrule
                           \multicolumn{6}{c}{\dataset{} Search Test Queries (Success@5)}                                        \\ \midrule
\scalebox{.85}{\textbf{Writing}}                    & 74.7                         & 60.3                         & 74.4                    & 78.0                      & 77.1                         & \textbf{80.1}             \\
\scalebox{.85}{\textbf{Recreation}}              & 68.5                         & 56.5                            & 64.7                    & 72.1                      & 69.0                         & \textbf{72.3}             \\
\scalebox{.85}{\textbf{Science}}                  & 53.6                         & 32.7                           & 53.6                    & 55.3                      & 55.4                         & \textbf{56.7}             \\
\scalebox{.85}{\textbf{Technology}}                 & 61.9                         & 41.8                         & 59.6                    & 63.4                      & 62.4                         & \textbf{66.1}             \\
\scalebox{.85}{\textbf{Lifestyle}}                  & 80.2                         & 63.8                         & 82.3                    & 82.1                      & 82.3                         & \textbf{84.7}             \\
\scalebox{.85}{\textbf{Pooled}}                     & 67.3                         & 48.3                         & 66.4                    & 69.8                   & 68.9                         & \textbf{71.6}             \\ \midrule
                           \multicolumn{6}{c}{\dataset{} Forum Test Queries (Success@5)}                                         \\ \midrule
\scalebox{.85}{\textbf{Writing}}                    & 71.0                         & 64.0                         & 68.8                    & 71.5                      & 73.0                         & \textbf{76.3}             \\
\scalebox{.85}{\textbf{Recreation}}              & 65.6                         & 55.4                         & 63.8                       & 65.7                      & 67.1                         & \textbf{70.8}             \\
\scalebox{.85}{\textbf{Science}}                  & 41.8                         & 37.1                         & 36.5                      & 38.0                      & 43.7                         & \textbf{46.1}             \\
\scalebox{.85}{\textbf{Technology}}                 & 48.5                         & 39.4                         & 46.8                    & 47.3                     & 50.8                         & \textbf{53.6}             \\
\scalebox{.85}{\textbf{Lifestyle}}                  & 73.0                         & 60.6                         & 73.1                    & 73.7                      & 74.0                         & \textbf{76.9}             \\
\scalebox{.85}{\textbf{Pooled}}                     & 58.2                         & 47.2                         & 55.7                    & 57.7                    & 60.1                         & \textbf{63.4}             \\ \bottomrule
\end{tabular}
\caption{} %
\label{table:wikipedia_and_coffee_results}
\end{subtable}
\vspace{-3mm}
\caption{Zero-shot evaluation results. Sub-table (a) reports results on BEIR and sub-table (b) reports results on the Wikipedia Open QA and the test sets of the \dataset{} benchmark. On BEIR, we test \system{} and RocketQAv2 and copy the results for ANCE, TAS-B, and ColBERT from \citet{thakur2021beir}, for MoDIR and DPR-MSMARCO (DPR-M) from \citet{xin2021zero}, and for SPLADEv2 from \citet{formal2021spladev2}.}
\label{table:ood_results}
\vspace{-3mm}
\end{table*}

\subsection{In-Domain Retrieval Quality} \label{sec:in_domain}

Similar to related work, we train for IR tasks on MS MARCO Passage Ranking~\cite{nguyen2016ms}. Within the training domain, our development-set results are shown in Table~\ref{table:in_domain_results}, comparing \system{} with vanilla ColBERT as well as state-of-the-art single-vector systems.

While ColBERT outperforms single-vector systems like RepBERT, ANCE, and even TAS-B, improvements in supervision such as distillation from cross-encoders enable systems like SPLADEv2, PAIR, and RocketQAv2 to achieve higher quality than vanilla ColBERT. These supervision gains challenge the value of fine-grained late interaction, and it is not inherently clear whether the stronger inductive biases of ColBERT-like models permit it to accept similar gains under distillation, especially when using compressed representations. Despite this, we find that with denoised supervision and residual compression,  \system{} achieves the highest quality across all systems. As we discuss in~\secref{sec:efficiency}, it exhibits space footprint competitive with these single-vector models and much lower than vanilla ColBERT.

Besides the official dev set, we evaluated \system{}, SPLADEv2, and RocketQAv2 on the ``Local Eval'' test set described by \citet{khattab2020colbert} for MS MARCO, which consists of 5000 queries disjoint with the training and the official dev sets. These queries are obtained from labeled 50k queries that are provided in the official MS MARCO Passage Ranking task as additional validation data.\footnote{These are sampled from delta between \texttt{qrels.dev.tsv} and \texttt{qrels.dev.small.tsv} on \url{https://microsoft.github.io/msmarco/Datasets}. We refer to \citet{khattab2020colbert} for details. All our query IDs will be made public to aid reproducibility.} On this test set, \system{} obtains 40.8\% MRR@10, considerably outperforming the baselines, including RocketQAv2 which makes use of document titles in addition to the passage text unlike the other systems.

\subsection{Out-of-Domain Retrieval Quality} \label{sec:out_of_domain}

Next, we evaluate \system{} outside the training domain using BEIR~\cite{thakur2021beir}, Wikipedia Open QA retrieval as in~\citet{khattab2021relevance}, and \dataset{}. We compare against a wide range of recent and state-of-the-art retrieval systems from the literature. %

\textbf{BEIR.} We start with BEIR, reporting the quality of models that do not incorporate distillation from cross-encoders, namely, ColBERT~\cite{khattab2020colbert}, DPR-MARCO~\cite{xin2021zero}, ANCE~\cite{xiong2020approximate},
and
MoDIR~\cite{xin2021zero}, as well as models that do utilize distillation, namely, TAS-B~\cite{hofstatter2021efficiently}, SPLADEv2~\cite{formal2021spladev2}, and also RocketQAv2, which we test ourselves using the official checkpoint trained on MS MARCO. We divide the table into ``search'' (i.e., natural queries and questions) and ``semantic relatednes'' (e.g., citation-relatedness and claim verification) tasks to reflect the nature of queries in each dataset.\footnote{Following \citet{formal2021spladev2}, we conduct our evaluationg using the publicly-available datasets in BEIR. Refer to~\secref{appendix:beir} for details.}

Table~\ref{table:beir_results} reports results with the official nDCG@10 metric. Among the models without distillation, we see that the vanilla ColBERT model outperforms the single-vector systems DPR, ANCE, and MoDIR across all but three tasks. ColBERT often outpaces all three systems by large margins and, in fact, outperforms the TAS-B model, which utilizes distillation, on most datasets. Shifting our attention to models with distillation, we see a similar pattern: while distillation-based models are generally stronger than their vanilla counterparts, the models that decompose scoring into term-level interactions, \system{} and SPLADEv2, are almost always the strongest. %

Looking more closely into the comparison between SPLADEv2 and \system{}, we see that \system{} has an advantage on six benchmarks and ties SPLADEv2 on two, with the largest improvements attained on NQ, TREC-COVID, and FiQA-2018, all of which feature natural search queries. On the other hand, SPLADEv2 has the lead on five benchmarks, displaying the largest gains on Climate-FEVER (C-FEVER) and HotPotQA. In C-FEVER, the input queries are sentences making climate-related claims and, as a result, do not reflect the typical characteristics of search queries. In HotPotQA, queries are written by crowdworkers who have access to the target passages. This is known to lead to artificial lexical bias~\cite{lee2019latent}, where crowdworkers copy terms from the passages into their questions as in the Open-SQuAD benchmark. %

\textbf{Wikipedia Open QA.} As a further test of out-of-domain generalization, we evaluate the MS MARCO-trained \system{}, SPLADEv2, and vanilla ColBERT on retrieval for open-domain question answering, similar to the out-of-domain setting of \citet{khattab2021relevance}. We report Success@5 (sometimes referred to as Recall@5), which is the percentage of questions whose short answer string overlaps with one or more of the top-5 passages. For the queries, we use the development set questions of the open-domain versions \cite{lee2019latent,karpukhin2020dense} of Natural Questions (NQ; \citealt{kwiatkowski2019natural}), TriviaQA (TQ; \citealt{joshi2017triviaqa}), and SQuAD~\cite{rajpurkar2016squad} datasets in Table~\ref{table:wikipedia_and_coffee_results}. As a baseline, we include the BM25~\cite{robertson1995okapi} results using the Anserini~\cite{yang2018anserini} toolkit. We observe that \system{} outperforms BM25, vanilla ColBERT, and SPLADEv2 across the three query sets, with improvements of up to 4.6 points over SPLADEv2.

\textbf{\dataset{}.} Next, we analyze performance on the \dataset{} test benchmark, which focuses on natural queries over long-tail topics and exhibits a different annotation pattern to the datasets in the previous OOD evaluations. In particular, \dataset{} uses automatic Google rankings (for the ``search'' queries) and organic StackExchange question–answer pairs (for ``forum'' queries), complimenting the pooling-based annotation of datasets like TREC-COVID (in BEIR) and the answer overlap metrics of Open-QA retrieval. %
We report Success@5 for each corpus on both search queries and forum queries.

Overall, we see that ANCE and vanilla ColBERT outperform BM25 on all topics, and that the three methods using distillation are generally the strongest. Similar to the Wikipedia-OpenQA results, we find that \system{} outperforms the baselines across all topics for both query types, improving upon SPLADEv2 and RocketQAv2 by up to 3.7 and 8.1 points, respectively. Considering the baselines, we observe that while RocketQAv2 tends to have a slight advantage over SPLADEv2 on the ``search'' queries, SPLADEv2 is considerably more effective on the ``forum'' tests. We hypothesize that the search queries, obtained from Google (through GooAQ) are more similar to MS MARCO than the forum queries and, as a result, the latter stresses generalization more heavily, rewarding term-decomposed models like SPLADEv2 and \system{}.

\subsection{Efficiency} \label{sec:efficiency}

\system{}'s residual compression approach significantly reduces index sizes compared to vanilla ColBERT. Whereas ColBERT requires 154 GiB to store the index for MS MARCO, \system{} only requires 16 GiB or 25 GiB when compressing embeddings to 1 or 2 bit(s) per dimension, respectively, resulting in compression ratios of 6--10$\times$. This storage figure includes 4.5 GiB for storing the inverted list.

This matches the storage for a typical single-vector model on MS MARCO, with 4-byte lossless floating-point storage for one 768-dimensional vector for each of the 9M passages amounting to a little over 25 GiBs. In practice, the storage for a single-vector model could be even larger when using a nearest-neighbor index like HNSW for fast search. Conversely, single-vector representations could be themselves compressed very aggressively~\cite{zhan2021jointly,zhan2021learning}, though often exacerbating the loss in quality relative to late interaction methods like \system{}. %

We discuss the impact of our compression method on search quality in Appendix~\ref{appendix:compression} and present query latency results on the order of 50--250 milliseconds per query in Appendix~\ref{appendix:latency}.

\section{Conclusion} \label{sec:conclusion}

We introduced \system{}, a retriever that advances the quality and space efficiency of multi-vector representations. We hypothesized that cluster centroids capture context-aware semantics of the token-level representations and proposed a residual representation that leverages these patterns to dramatically reduce the footprint of multi-vector systems \textit{off-the-shelf}. We then explored improved supervision for multi-vector retrieval and found that their quality improves considerably upon distillation from a cross-encoder system. The proposed \system{} considerably outperforms existing retrievers in within-domain and out-of-domain evaluations, which we conducted extensively across 28 datasets, establishing state-of-the-art quality while exhibiting competitive space footprint.

\section*{Acknowledgements}

This research was supported in part by affiliate members and other supporters of the Stanford DAWN project---Ant Financial, Facebook, Google, and VMware---as well as Cisco, SAP, Virtusa, and the NSF under CAREER grant CNS-1651570. Any opinions, findings, and conclusions or recommendations expressed in this material are those of the authors and do not necessarily reflect the views of the National Science Foundation.

\section*{Broader Impact \& Ethical Considerations}

This work is primarily an effort toward retrieval models that generalize better while performing reasonably efficiently in terms of space consumption. Strong out-of-the-box generalization to small domain-specific applications can serve many users in practice, particularly where training data is not available. Moreover, retrieval holds significant promise for many downstream NLP tasks, as it can help make language models smaller and thus more efficient (i.e., by decoupling knowledge from computation), more transparent (i.e., by allowing users to check the sources the model relied on when making a claim or prediction), and easier to update (i.e., by allowing developers to replace or add documents to the corpus without retraining the model) \cite{guu2020realm,borgeaud2021improving,khattab2021baleen}. Nonetheless, such work poses risks in terms of misuse, particularly toward misinformation, as retrieval can surface results that are relevant yet inaccurate, depending on the contents of a corpus. Moreover, generalization from training on a large-scale dataset can propagate the biases of that dataset well beyond its typical reach to new domains and applications.

While our contributions have made ColBERT's late interaction more efficient at storage costs, large-scale distillation with hard negatives increases system complexity and accordingly increases training cost, when compared with the straightforward training paradigm of the original ColBERT model. While \system{} is efficient in terms of latency and storage at inference time, we suspect that under extreme resource constraints, simpler model designs like SPLADEv2 or RocketQAv2 could lend themselves to easier-to-optimize environments. We leave low-level systems optimizations of all systems to future work. Another worthwhile dimension for future exploration of tradeoffs is re-ranking architectures over various systems with cross-encoders, which are known to be expensive yet precise due to their highly expressive capacity.

\section*{Research Limitations}
\label{appendix:limitations}

While we evaluate \system{} on a wide range of tests, all of our benchmarks are in English and, in line with related work, our out-of-domain tests evaluate models that are trained on MS MARCO. We expect our approach to work effectively for other languages and when all models are trained using other, smaller training set (e.g., NaturalQuestions), but we leave such tests to future work.

We have observed consistent gains for \system{} against existing state-of-the-art systems across many diverse settings. Despite this, almost all IR datasets contain false negatives (i.e., relevant but unlabeled passages) and thus some caution is needed in interpreting any individual result. Nonetheless, we intentionally sought out benchmarks with dissimilar annotation biases: for instance, TREC-COVID (in BEIR) annotates the pool of documents retrieved by the systems submitted at the time of the competition, \dataset{} uses automatic Google rankings (for ``search'' queries) and StackExchange question--answer pairs (for ``forum'' queries), and the Open-QA tests rely on passage-answer overlap for factoid questions. \system{} performed well in all of these settings. We discuss other issues pertinent to \dataset{} in Appendix~\secref{section:appendix_lotte}.

We have compared with a wide range of strong baselines---including sparse retrieval and single-vector models---and found reliable patterns across tests. However, we caution that empirical trends can change as innovations are introduced to each of these families of models and that it can be difficult to ensure exact apple-to-apple comparisons across families of models, since each of them calls for different sophisticated tuning strategies. We thus primarily used results and models from the rich recent literature on these problems, with models like RocketQAv2 and SPLADEv2.

On the representational side, we focus on reducing the storage cost using residual compression, achieving strong gains in reducing footprint while largely preserving quality. Nonetheless, we have not exhausted the space of more sophisticated optimizations possible, and we would expect more sophisticated forms of residual compression and composing our approach with dropping tokens \cite{zhou-devlin-2021-multi} to open up possibilities for further reductions in space footprint.

\bibliography{anthology,bibliography}

\begin{thebibliography}{77}
\expandafter\ifx\csname natexlab\endcsname\relax\def\natexlab#1{#1}\fi

\bibitem[{sta()}]{stackexchange}

\newblock \href {https://archive.org/details/stackexchange} {{S}tack {E}xchange
  {D}ata {D}ump}.

\bibitem[{Ai et~al.(2017)Ai, Yu, Wu, He, and Guan}]{ai2017optimized}
Liefu Ai, Junqing Yu, Zebin Wu, Yunfeng He, and Tao Guan. 2017.
\newblock {O}ptimized {R}esidual {V}ector {Q}uantization for {E}fficient
  {A}pproximate {N}earest {N}eighbor {S}earch.
\newblock \emph{Multimedia Systems}, 23(2):169--181.

\bibitem[{Auer et~al.(2007)Auer, Bizer, Kobilarov, Lehmann, Cyganiak, and
  Ives}]{auer2007dbpedia}
S{\"o}ren Auer, Christian Bizer, Georgi Kobilarov, Jens Lehmann, Richard
  Cyganiak, and Zachary Ives. 2007.
\newblock {DB}pedia: A {N}ucleus for a {W}eb of {O}pen {D}ata.
\newblock In \emph{The semantic web}, pages 722--735. Springer.

\bibitem[{Barnes et~al.(1996)Barnes, Rizvi, and Nasrabadi}]{barnes1996advances}
Christopher~F Barnes, Syed~A Rizvi, and Nasser~M Nasrabadi. 1996.
\newblock {A}dvances in {R}esidual {V}ector {Q}uantization: A {R}eview.
\newblock \emph{IEEE transactions on image processing}, 5(2):226--262.

\bibitem[{Bondarenko et~al.(2020)Bondarenko, Fr{\"o}be, Beloucif, Gienapp,
  Ajjour, Panchenko, Biemann, Stein, Wachsmuth, Potthast
  et~al.}]{bondarenko2020overview}
Alexander Bondarenko, Maik Fr{\"o}be, Meriem Beloucif, Lukas Gienapp, Yamen
  Ajjour, Alexander Panchenko, Chris Biemann, Benno Stein, Henning Wachsmuth,
  Martin Potthast, et~al. 2020.
\newblock {O}verview of touch{\'e} 2020: {A}rgument {R}etrieval.
\newblock In \emph{International Conference of the Cross-Language Evaluation
  Forum for European Languages}, pages 384--395. Springer.

\bibitem[{Borgeaud et~al.(2021)Borgeaud, Mensch, Hoffmann, Cai, Rutherford,
  Millican, Driessche, Lespiau, Damoc, Clark et~al.}]{borgeaud2021improving}
Sebastian Borgeaud, Arthur Mensch, Jordan Hoffmann, Trevor Cai, Eliza
  Rutherford, Katie Millican, George van~den Driessche, Jean-Baptiste Lespiau,
  Bogdan Damoc, Aidan Clark, et~al. 2021.
\newblock \href {https://arxiv.org/abs/2112.04426} {Improving language models
  by retrieving from trillions of tokens}.
\newblock \emph{arXiv preprint arXiv:2112.04426}.

\bibitem[{Boteva et~al.(2016)Boteva, Gholipour, Sokolov, and
  Riezler}]{boteva2016full}
Vera Boteva, Demian Gholipour, Artem Sokolov, and Stefan Riezler. 2016.
\newblock A {F}ull-text {L}earning to {R}ank {D}ataset for {M}edical
  {I}nformation {R}etrieval.
\newblock In \emph{European Conference on Information Retrieval}, pages
  716--722. Springer.

\bibitem[{Chen et~al.(2018)Chen, Choi, Brand, Agrawal, Zhang, and
  Gopalakrishnan}]{chen2018adacomp}
Chia{-}Yu Chen, Jungwook Choi, Daniel Brand, Ankur Agrawal, Wei Zhang, and
  Kailash Gopalakrishnan. 2018.
\newblock \href
  {https://www.aaai.org/ocs/index.php/AAAI/AAAI18/paper/view/16859} {Adacomp :
  Adaptive residual gradient compression for data-parallel distributed
  training}.
\newblock In \emph{Proceedings of the Thirty-Second {AAAI} Conference on
  Artificial Intelligence, (AAAI-18), the 30th innovative Applications of
  Artificial Intelligence (IAAI-18), and the 8th {AAAI} Symposium on
  Educational Advances in Artificial Intelligence (EAAI-18), New Orleans,
  Louisiana, USA, February 2-7, 2018}, pages 2827--2835. {AAAI} Press.

\bibitem[{Cohan et~al.(2020)Cohan, Feldman, Beltagy, Downey, and
  Weld}]{cohan2020specter}
Arman Cohan, Sergey Feldman, Iz~Beltagy, Doug Downey, and Daniel Weld. 2020.
\newblock \href {https://doi.org/10.18653/v1/2020.acl-main.207} {{SPECTER}:
  Document-level representation learning using citation-informed transformers}.
\newblock In \emph{Proceedings of the 58th Annual Meeting of the Association
  for Computational Linguistics}, pages 2270--2282, Online. Association for
  Computational Linguistics.

\bibitem[{Cohen et~al.(2021)Cohen, Portnoy, Fetahu, and Ingber}]{cohen2021sdr}
Nachshon Cohen, Amit Portnoy, Besnik Fetahu, and Amir Ingber. 2021.
\newblock \href {https://arxiv.org/abs/2110.02065} {{SDR}: {E}fficient {N}eural
  {R}e-ranking using {S}uccinct {D}ocument {R}epresentation}.
\newblock \emph{arXiv preprint arXiv:2110.02065}.

\bibitem[{Dai and Callan(2020)}]{dai2020context}
Zhuyun Dai and Jamie Callan. 2020.
\newblock \href {https://doi.org/10.1145/3397271.3401204} {Context-aware term
  weighting for first stage passage retrieval}.
\newblock In \emph{Proceedings of the 43rd International {ACM} {SIGIR}
  conference on research and development in Information Retrieval, {SIGIR}
  2020, Virtual Event, China, July 25-30, 2020}, pages 1533--1536. {ACM}.

\bibitem[{Devlin et~al.(2019)Devlin, Chang, Lee, and
  Toutanova}]{devlin-etal-2019-bert}
Jacob Devlin, Ming-Wei Chang, Kenton Lee, and Kristina Toutanova. 2019.
\newblock \href {https://doi.org/10.18653/v1/N19-1423} {{BERT}: Pre-training of
  deep bidirectional transformers for language understanding}.
\newblock In \emph{Proceedings of the 2019 Conference of the North {A}merican
  Chapter of the Association for Computational Linguistics: Human Language
  Technologies, Volume 1 (Long and Short Papers)}, pages 4171--4186,
  Minneapolis, Minnesota. Association for Computational Linguistics.

\bibitem[{Diggelmann et~al.(2020)Diggelmann, Boyd-Graber, Bulian, Ciaramita,
  and Leippold}]{diggelmann2020climate}
Thomas Diggelmann, Jordan Boyd-Graber, Jannis Bulian, Massimiliano Ciaramita,
  and Markus Leippold. 2020.
\newblock \href {https://arxiv.org/abs/2012.00614} {{CLIMATE}-{FEVER}: A
  {D}ataset for {V}erification of {R}eal-{W}orld {C}limate {C}laims}.
\newblock \emph{arXiv preprint arXiv:2012.00614}.

\bibitem[{Formal et~al.(2021{\natexlab{a}})Formal, Lassance, Piwowarski, and
  Clinchant}]{formal2021spladev2}
Thibault Formal, Carlos Lassance, Benjamin Piwowarski, and St{\'e}phane
  Clinchant. 2021{\natexlab{a}}.
\newblock \href {https://arxiv.org/abs/2109.10086} {{SPLADE} v2: {S}parse
  {L}exical and {E}xpansion {M}odel for {I}nformation {R}etrieval}.
\newblock \emph{arXiv preprint arXiv:2109.10086}.

\bibitem[{Formal et~al.(2021{\natexlab{b}})Formal, Piwowarski, and
  Clinchant}]{formal2021splade}
Thibault Formal, Benjamin Piwowarski, and St{\'e}phane Clinchant.
  2021{\natexlab{b}}.
\newblock {SPLADE}: {S}parse {L}exical and {E}xpansion {M}odel for {F}irst
  {S}tage {R}anking.
\newblock In \emph{Proceedings of the 44th International ACM SIGIR Conference
  on Research and Development in Information Retrieval}, pages 2288--2292.

\bibitem[{Gao and Callan(2021)}]{gao2021unsupervised}
Luyu Gao and Jamie Callan. 2021.
\newblock \href {https://arxiv.org/abs/2108.05540} {Unsupervised corpus aware
  language model pre-training for dense passage retrieval}.
\newblock \emph{arXiv preprint arXiv:2108.05540}.

\bibitem[{Gao et~al.(2020)Gao, Dai, and Callan}]{gao2020modularized}
Luyu Gao, Zhuyun Dai, and Jamie Callan. 2020.
\newblock \href {https://doi.org/10.18653/v1/2020.emnlp-main.342} {Modularized
  transfomer-based ranking framework}.
\newblock In \emph{Proceedings of the 2020 Conference on Empirical Methods in
  Natural Language Processing (EMNLP)}, pages 4180--4190, Online. Association
  for Computational Linguistics.

\bibitem[{Gao et~al.(2021)Gao, Dai, and Callan}]{gao2021coil}
Luyu Gao, Zhuyun Dai, and Jamie Callan. 2021.
\newblock \href {https://doi.org/10.18653/v1/2021.naacl-main.241} {{COIL}:
  Revisit exact lexical match in information retrieval with contextualized
  inverted list}.
\newblock In \emph{Proceedings of the 2021 Conference of the North American
  Chapter of the Association for Computational Linguistics: Human Language
  Technologies}, pages 3030--3042, Online. Association for Computational
  Linguistics.

\bibitem[{Gray(1984)}]{gray1984vector}
Robert Gray. 1984.
\newblock Vector quantization.
\newblock \emph{IEEE Assp Magazine}, 1(2):4--29.

\bibitem[{Guu et~al.(2020)Guu, Lee, Tung, Pasupat, and Chang}]{guu2020realm}
Kelvin Guu, Kenton Lee, Zora Tung, Panupong Pasupat, and Ming-Wei Chang. 2020.
\newblock \href {https://arxiv.org/abs/2002.08909} {Realm: Retrieval-augmented
  language model pre-training}.
\newblock \emph{arXiv preprint arXiv:2002.08909}.

\bibitem[{Hofst{\"a}tter et~al.(2020)Hofst{\"a}tter, Althammer, Schr{\"o}der,
  Sertkan, and Hanbury}]{hofstatter2020improving}
Sebastian Hofst{\"a}tter, Sophia Althammer, Michael Schr{\"o}der, Mete Sertkan,
  and Allan Hanbury. 2020.
\newblock \href {https://arxiv.org/abs/2010.02666} {{I}mproving {E}fficient
  {N}eural {R}anking {M}odels with {C}ross-{A}rchitecture {K}nowledge
  {D}istillation}.
\newblock \emph{arXiv preprint arXiv:2010.02666}.

\bibitem[{Hofst{\"a}tter et~al.(2021)Hofst{\"a}tter, Lin, Yang, Lin, and
  Hanbury}]{hofstatter2021efficiently}
Sebastian Hofst{\"a}tter, Sheng-Chieh Lin, Jheng-Hong Yang, Jimmy Lin, and
  Allan Hanbury. 2021.
\newblock \href {https://arxiv.org/abs/2104.06967} {{E}fficiently {T}eaching an
  {E}ffective {D}ense {R}etriever with {B}alanced {T}opic {A}ware {S}ampling}.
\newblock \emph{arXiv preprint arXiv:2104.06967}.

\bibitem[{Humeau et~al.(2020)Humeau, Shuster, Lachaux, and
  Weston}]{humeau2020polyencoders}
Samuel Humeau, Kurt Shuster, Marie{-}Anne Lachaux, and Jason Weston. 2020.
\newblock \href {https://openreview.net/forum?id=SkxgnnNFvH} {Poly-encoders:
  Architectures and pre-training strategies for fast and accurate
  multi-sentence scoring}.
\newblock In \emph{8th International Conference on Learning Representations,
  {ICLR} 2020, Addis Ababa, Ethiopia, April 26-30, 2020}. OpenReview.net.

\bibitem[{Izacard et~al.(2020)Izacard, Petroni, Hosseini, De~Cao, Riedel, and
  Grave}]{izacard2020memory}
Gautier Izacard, Fabio Petroni, Lucas Hosseini, Nicola De~Cao, Sebastian
  Riedel, and Edouard Grave. 2020.
\newblock \href {https://arxiv.org/abs/2012.15156} {A memory efficient baseline
  for open domain question answering}.
\newblock \emph{arXiv preprint arXiv:2012.15156}.

\bibitem[{Jegou et~al.(2010)Jegou, Douze, and Schmid}]{jegou2010product}
Herve Jegou, Matthijs Douze, and Cordelia Schmid. 2010.
\newblock Product quantization for nearest neighbor search.
\newblock \emph{IEEE transactions on pattern analysis and machine
  intelligence}, 33(1):117--128.

\bibitem[{Jiang et~al.(2020)Jiang, Bordia, Zhong, Dognin, Singh, and
  Bansal}]{jiang2020hover}
Yichen Jiang, Shikha Bordia, Zheng Zhong, Charles Dognin, Maneesh Singh, and
  Mohit Bansal. 2020.
\newblock \href {https://doi.org/10.18653/v1/2020.findings-emnlp.309}
  {{H}o{V}er: A dataset for many-hop fact extraction and claim verification}.
\newblock In \emph{Findings of the Association for Computational Linguistics:
  EMNLP 2020}, pages 3441--3460, Online. Association for Computational
  Linguistics.

\bibitem[{Johnson et~al.(2019)Johnson, Douze, and
  J{\'e}gou}]{johnson2019billion}
Jeff Johnson, Matthijs Douze, and Herv{\'e} J{\'e}gou. 2019.
\newblock Billion-scale similarity search with gpus.
\newblock \emph{IEEE Transactions on Big Data}.

\bibitem[{Joshi et~al.(2017)Joshi, Choi, Weld, and
  Zettlemoyer}]{joshi2017triviaqa}
Mandar Joshi, Eunsol Choi, Daniel Weld, and Luke Zettlemoyer. 2017.
\newblock \href {https://doi.org/10.18653/v1/P17-1147} {{T}rivia{QA}: A large
  scale distantly supervised challenge dataset for reading comprehension}.
\newblock In \emph{Proceedings of the 55th Annual Meeting of the Association
  for Computational Linguistics (Volume 1: Long Papers)}, pages 1601--1611,
  Vancouver, Canada. Association for Computational Linguistics.

\bibitem[{Karpukhin et~al.(2020)Karpukhin, Oguz, Min, Lewis, Wu, Edunov, Chen,
  and Yih}]{karpukhin2020dense}
Vladimir Karpukhin, Barlas Oguz, Sewon Min, Patrick Lewis, Ledell Wu, Sergey
  Edunov, Danqi Chen, and Wen-tau Yih. 2020.
\newblock \href {https://doi.org/10.18653/v1/2020.emnlp-main.550} {Dense
  passage retrieval for open-domain question answering}.
\newblock In \emph{Proceedings of the 2020 Conference on Empirical Methods in
  Natural Language Processing (EMNLP)}, pages 6769--6781, Online. Association
  for Computational Linguistics.

\bibitem[{Khashabi et~al.(2021)Khashabi, Ng, Khot, Sabharwal, Hajishirzi, and
  Callison-Burch}]{khashabi2021gooaq}
Daniel Khashabi, Amos Ng, Tushar Khot, Ashish Sabharwal, Hannaneh Hajishirzi,
  and Chris Callison-Burch. 2021.
\newblock \href {https://arxiv.org/abs/2104.08727} {{G}oo{AQ}: {O}pen
  {Q}uestion {A}nswering with {D}iverse {A}nswer {T}ypes}.
\newblock \emph{arXiv preprint arXiv:2104.08727}.

\bibitem[{Khattab et~al.(2021{\natexlab{a}})Khattab, Potts, and
  Zaharia}]{khattab2021baleen}
Omar Khattab, Christopher Potts, and Matei Zaharia. 2021{\natexlab{a}}.
\newblock {B}aleen: {R}obust {M}ulti-{H}op {R}easoning at {S}cale via
  {C}ondensed {R}etrieval.
\newblock In \emph{Thirty-Fifth Conference on Neural Information Processing
  Systems}.

\bibitem[{Khattab et~al.(2021{\natexlab{b}})Khattab, Potts, and
  Zaharia}]{khattab2021relevance}
Omar Khattab, Christopher Potts, and Matei Zaharia. 2021{\natexlab{b}}.
\newblock Relevance-guided supervision for openqa with {ColBERT}.
\newblock \emph{Transactions of the Association for Computational Linguistics},
  9:929--944.

\bibitem[{Khattab and Zaharia(2020)}]{khattab2020colbert}
Omar Khattab and Matei Zaharia. 2020.
\newblock \href {https://doi.org/10.1145/3397271.3401075} {Colbert: Efficient
  and effective passage search via contextualized late interaction over
  {BERT}}.
\newblock In \emph{Proceedings of the 43rd International {ACM} {SIGIR}
  conference on research and development in Information Retrieval, {SIGIR}
  2020, Virtual Event, China, July 25-30, 2020}, pages 39--48. {ACM}.

\bibitem[{Kwiatkowski et~al.(2019)Kwiatkowski, Palomaki, Redfield, Collins,
  Parikh, Alberti, Epstein, Polosukhin, Devlin, Lee, Toutanova, Jones, Kelcey,
  Chang, Dai, Uszkoreit, Le, and Petrov}]{kwiatkowski2019natural}
Tom Kwiatkowski, Jennimaria Palomaki, Olivia Redfield, Michael Collins, Ankur
  Parikh, Chris Alberti, Danielle Epstein, Illia Polosukhin, Jacob Devlin,
  Kenton Lee, Kristina Toutanova, Llion Jones, Matthew Kelcey, Ming-Wei Chang,
  Andrew~M. Dai, Jakob Uszkoreit, Quoc Le, and Slav Petrov. 2019.
\newblock \href {https://doi.org/10.1162/tacl_a_00276} {Natural questions: A
  benchmark for question answering research}.
\newblock \emph{Transactions of the Association for Computational Linguistics},
  7:452--466.

\bibitem[{Lee et~al.(2021{\natexlab{a}})Lee, Sung, Kang, and
  Chen}]{lee-etal-2021-learning-dense}
Jinhyuk Lee, Mujeen Sung, Jaewoo Kang, and Danqi Chen. 2021{\natexlab{a}}.
\newblock \href {https://doi.org/10.18653/v1/2021.acl-long.518} {Learning dense
  representations of phrases at scale}.
\newblock In \emph{Proceedings of the 59th Annual Meeting of the Association
  for Computational Linguistics and the 11th International Joint Conference on
  Natural Language Processing (Volume 1: Long Papers)}, pages 6634--6647,
  Online. Association for Computational Linguistics.

\bibitem[{Lee et~al.(2021{\natexlab{b}})Lee, Wettig, and Chen}]{lee2021phrase}
Jinhyuk Lee, Alexander Wettig, and Danqi Chen. 2021{\natexlab{b}}.
\newblock \href {https://arxiv.org/abs/2109.08133} {Phrase retrieval learns
  passage retrieval, too}.
\newblock \emph{arXiv preprint arXiv:2109.08133}.

\bibitem[{Lee et~al.(2019)Lee, Chang, and Toutanova}]{lee2019latent}
Kenton Lee, Ming-Wei Chang, and Kristina Toutanova. 2019.
\newblock \href {https://doi.org/10.18653/v1/P19-1612} {Latent retrieval for
  weakly supervised open domain question answering}.
\newblock In \emph{Proceedings of the 57th Annual Meeting of the Association
  for Computational Linguistics}, pages 6086--6096, Florence, Italy.
  Association for Computational Linguistics.

\bibitem[{Li et~al.(2021{\natexlab{a}})Li, Ding, Liu, Zhang, and
  Guo}]{li2021trq}
Yue Li, Wenrui Ding, Chunlei Liu, Baochang Zhang, and Guodong Guo.
  2021{\natexlab{a}}.
\newblock {TRQ}: {T}ernary {N}eural {N}etworks {W}ith {R}esidual
  {Q}uantization.
\newblock In \emph{Proceedings of the AAAI Conference on Artificial
  Intelligence}, volume~35, pages 8538--8546.

\bibitem[{Li et~al.(2021{\natexlab{b}})Li, Ni, Li, Yang, Zhang, and
  Gao}]{li2021residual}
Zefan Li, Bingbing Ni, Teng Li, Xiaokang Yang, Wenjun Zhang, and Wen Gao.
  2021{\natexlab{b}}.
\newblock {R}esidual {Q}uantization for {L}ow {B}it-width {N}eural {N}etworks.
\newblock \emph{IEEE Transactions on Multimedia}.

\bibitem[{Lin and Ma(2021)}]{lin2021few}
Jimmy Lin and Xueguang Ma. 2021.
\newblock \href {https://arxiv.org/abs/2106.14807} {A {F}ew {B}rief {N}otes on
  {D}eep{I}mpact, {COIL}, and a {C}onceptual {F}ramework for {I}nformation
  {R}etrieval {T}echniques}.
\newblock \emph{arXiv preprint arXiv:2106.14807}.

\bibitem[{Lin et~al.(2020)Lin, Yang, and Lin}]{lin2020distilling}
Sheng-Chieh Lin, Jheng-Hong Yang, and Jimmy Lin. 2020.
\newblock \href {https://arxiv.org/abs/2010.11386} {{D}istilling {D}ense
  {R}epresentations for {R}anking using {T}ightly-{C}oupled {T}eachers}.
\newblock \emph{arXiv preprint arXiv:2010.11386}.

\bibitem[{Liu et~al.(2020)Liu, Li, Tang, and Yan}]{liu2020double}
Xiaorui Liu, Yao Li, Jiliang Tang, and Ming Yan. 2020.
\newblock \href {http://proceedings.mlr.press/v108/liu20a.html} {A double
  residual compression algorithm for efficient distributed learning}.
\newblock In \emph{The 23rd International Conference on Artificial Intelligence
  and Statistics, {AISTATS} 2020, 26-28 August 2020, Online [Palermo, Sicily,
  Italy]}, volume 108 of \emph{Proceedings of Machine Learning Research}, pages
  133--143. {PMLR}.

\bibitem[{Luan et~al.(2021)Luan, Eisenstein, Toutanova, and
  Collins}]{luan2021sparse}
Yi~Luan, Jacob Eisenstein, Kristina Toutanova, and Michael Collins. 2021.
\newblock {S}parse, {D}ense, and {A}ttentional {R}epresentations for {T}ext
  {R}etrieval.
\newblock \emph{Transactions of the Association for Computational Linguistics},
  9:329--345.

\bibitem[{MacAvaney et~al.(2020)MacAvaney, Nardini, Perego, Tonellotto,
  Goharian, and Frieder}]{macavaney2020efficient}
Sean MacAvaney, Franco~Maria Nardini, Raffaele Perego, Nicola Tonellotto, Nazli
  Goharian, and Ophir Frieder. 2020.
\newblock \href {https://doi.org/10.1145/3397271.3401093} {Efficient document
  re-ranking for transformers by precomputing term representations}.
\newblock In \emph{Proceedings of the 43rd International {ACM} {SIGIR}
  conference on research and development in Information Retrieval, {SIGIR}
  2020, Virtual Event, China, July 25-30, 2020}, pages 49--58. {ACM}.

\bibitem[{Macdonald and Tonellotto(2021)}]{macdonald2021approximate}
Craig Macdonald and Nicola Tonellotto. 2021.
\newblock On approximate nearest neighbour selection for multi-stage dense
  retrieval.
\newblock In \emph{Proceedings of the 30th ACM International Conference on
  Information \& Knowledge Management}, pages 3318--3322.

\bibitem[{Maia et~al.(2018)Maia, Handschuh, Freitas, Davis, McDermott, Zarrouk,
  and Balahur}]{maia2018fiqa}
Macedo Maia, Siegfried Handschuh, Andr{\'e} Freitas, Brian Davis, Ross
  McDermott, Manel Zarrouk, and Alexandra Balahur. 2018.
\newblock {WWW}'18 {O}pen {C}hallenge: {F}inancial {O}pinion {M}ining and
  {Q}uestion {A}nswering.
\newblock In \emph{Companion Proceedings of the The Web Conference 2018}, pages
  1941--1942.

\bibitem[{Mallia et~al.(2021)Mallia, Khattab, Suel, and
  Tonellotto}]{mallia2021learning}
Antonio Mallia, Omar Khattab, Torsten Suel, and Nicola Tonellotto. 2021.
\newblock Learning passage impacts for inverted indexes.
\newblock In \emph{Proceedings of the 44th International ACM SIGIR Conference
  on Research and Development in Information Retrieval}, pages 1723--1727.

\bibitem[{Menon et~al.(2022)Menon, Jayasumana, Kim, Rawat, Reddi, and
  Kumar}]{menon2022in}
Aditya~Krishna Menon, Sadeep Jayasumana, Seungyeon Kim, Ankit~Singh Rawat,
  Sashank~J. Reddi, and Sanjiv Kumar. 2022.
\newblock \href {https://openreview.net/forum?id=bglU8l_Pq8Q} {In defense of
  dual-encoders for neural ranking}.

\bibitem[{Nguyen et~al.(2016)Nguyen, Rosenberg, Song, Gao, Tiwary, Majumder,
  and Deng}]{nguyen2016ms}
Tri Nguyen, Mir Rosenberg, Xia Song, Jianfeng Gao, Saurabh Tiwary, Rangan
  Majumder, and Li~Deng. 2016.
\newblock \href {https://arxiv.org/abs/1611.09268} {{MS MARCO}: A
  human-generated {MA}chine reading {CO}mprehension dataset}.
\newblock \emph{arXiv preprint arXiv:1611.09268}.

\bibitem[{Nogueira and Cho(2019)}]{nogueira2019passage}
Rodrigo Nogueira and Kyunghyun Cho. 2019.
\newblock \href {https://arxiv.org/abs/1901.04085} {{P}assage {R}e-ranking with
  {BERT}}.
\newblock \emph{arXiv preprint arXiv:1901.04085}.

\bibitem[{O{\u{g}}uz et~al.(2021)O{\u{g}}uz, Lakhotia, Gupta, Lewis, Karpukhin,
  Piktus, Chen, Riedel, Yih, Gupta et~al.}]{ouguz2021domain}
Barlas O{\u{g}}uz, Kushal Lakhotia, Anchit Gupta, Patrick Lewis, Vladimir
  Karpukhin, Aleksandra Piktus, Xilun Chen, Sebastian Riedel, Wen-tau Yih,
  Sonal Gupta, et~al. 2021.
\newblock \href {https://arxiv.org/abs/2107.13602} {{D}omain-matched
  {P}re-training {T}asks for {D}ense {R}etrieval}.
\newblock \emph{arXiv preprint arXiv:2107.13602}.

\bibitem[{Paranjape et~al.(2022)Paranjape, Khattab, Potts, Zaharia, and
  Manning}]{paranjape2021hindsight}
Ashwin Paranjape, Omar Khattab, Christopher Potts, Matei Zaharia, and
  Christopher~D Manning. 2022.
\newblock \href {https://openreview.net/forum?id=Vr_BTpw3wz} {Hindsight:
  Posterior-guided training of retrievers for improved open-ended generation}.
\newblock In \emph{International Conference on Learning Representations}.

\bibitem[{Qu et~al.(2021)Qu, Ding, Liu, Liu, Ren, Zhao, Dong, Wu, and
  Wang}]{qu2021rocketqa}
Yingqi Qu, Yuchen Ding, Jing Liu, Kai Liu, Ruiyang Ren, Wayne~Xin Zhao, Daxiang
  Dong, Hua Wu, and Haifeng Wang. 2021.
\newblock \href {https://doi.org/10.18653/v1/2021.naacl-main.466}
  {{R}ocket{QA}: An optimized training approach to dense passage retrieval for
  open-domain question answering}.
\newblock In \emph{Proceedings of the 2021 Conference of the North American
  Chapter of the Association for Computational Linguistics: Human Language
  Technologies}, pages 5835--5847, Online. Association for Computational
  Linguistics.

\bibitem[{Rajpurkar et~al.(2016)Rajpurkar, Zhang, Lopyrev, and
  Liang}]{rajpurkar2016squad}
Pranav Rajpurkar, Jian Zhang, Konstantin Lopyrev, and Percy Liang. 2016.
\newblock \href {https://doi.org/10.18653/v1/D16-1264} {{SQ}u{AD}: 100,000+
  questions for machine comprehension of text}.
\newblock In \emph{Proceedings of the 2016 Conference on Empirical Methods in
  Natural Language Processing}, pages 2383--2392, Austin, Texas. Association
  for Computational Linguistics.

\bibitem[{Ren et~al.(2021{\natexlab{a}})Ren, Lv, Qu, Liu, Zhao, She, Wu, Wang,
  and Wen}]{ren2021pair}
Ruiyang Ren, Shangwen Lv, Yingqi Qu, Jing Liu, Wayne~Xin Zhao, QiaoQiao She,
  Hua Wu, Haifeng Wang, and Ji-Rong Wen. 2021{\natexlab{a}}.
\newblock \href {https://doi.org/10.18653/v1/2021.findings-acl.191} {{PAIR}:
  Leveraging passage-centric similarity relation for improving dense passage
  retrieval}.
\newblock In \emph{Findings of the Association for Computational Linguistics:
  ACL-IJCNLP 2021}, pages 2173--2183, Online. Association for Computational
  Linguistics.

\bibitem[{Ren et~al.(2021{\natexlab{b}})Ren, Qu, Liu, Zhao, She, Wu, Wang, and
  Wen}]{ren2021rocketqav2}
Ruiyang Ren, Yingqi Qu, Jing Liu, Wayne~Xin Zhao, Qiaoqiao She, Hua Wu, Haifeng
  Wang, and Ji-Rong Wen. 2021{\natexlab{b}}.
\newblock \href {https://arxiv.org/abs/2110.07367} {{R}ocket{QA}v2: A {J}oint
  {T}raining {M}ethod for {D}ense {P}assage {R}etrieval and {P}assage
  {R}e-ranking}.
\newblock \emph{arXiv preprint arXiv:2110.07367}.

\bibitem[{Robertson et~al.(1995)Robertson, Walker, Jones, Hancock-Beaulieu,
  Gatford et~al.}]{robertson1995okapi}
Stephen~E Robertson, Steve Walker, Susan Jones, Micheline~M Hancock-Beaulieu,
  Mike Gatford, et~al. 1995.
\newblock Okapi at {TREC-3}.
\newblock \emph{NIST Special Publication}.

\bibitem[{Thakur et~al.(2021)Thakur, Reimers, R{\"u}ckl{\'e}, Srivastava, and
  Gurevych}]{thakur2021beir}
Nandan Thakur, Nils Reimers, Andreas R{\"u}ckl{\'e}, Abhishek Srivastava, and
  Iryna Gurevych. 2021.
\newblock \href {https://arxiv.org/abs/2104.08663} {{BEIR}: A {H}eterogenous
  {B}enchmark for {Z}ero-shot {E}valuation of {I}nformation {R}etrieval
  {M}odels}.
\newblock \emph{arXiv preprint arXiv:2104.08663}.

\bibitem[{Thorne et~al.(2018)Thorne, Vlachos, Christodoulopoulos, and
  Mittal}]{thorne2018fever}
James Thorne, Andreas Vlachos, Christos Christodoulopoulos, and Arpit Mittal.
  2018.
\newblock \href {https://doi.org/10.18653/v1/N18-1074} {{FEVER}: a large-scale
  dataset for fact extraction and {VER}ification}.
\newblock In \emph{Proceedings of the 2018 Conference of the North {A}merican
  Chapter of the Association for Computational Linguistics: Human Language
  Technologies, Volume 1 (Long Papers)}, pages 809--819, New Orleans,
  Louisiana. Association for Computational Linguistics.

\bibitem[{Voorhees et~al.(2021)Voorhees, Alam, Bedrick, Demner-Fushman, Hersh,
  Lo, Roberts, Soboroff, and Wang}]{voorhees2021trec}
Ellen Voorhees, Tasmeer Alam, Steven Bedrick, Dina Demner-Fushman, William~R
  Hersh, Kyle Lo, Kirk Roberts, Ian Soboroff, and Lucy~Lu Wang. 2021.
\newblock {TREC}-{COVID}: {C}onstructing a {P}andemic {I}nformation {R}etrieval
  {T}est {C}ollection.
\newblock In \emph{ACM SIGIR Forum}, volume~54, pages 1--12. ACM New York, NY,
  USA.

\bibitem[{Wachsmuth et~al.(2018)Wachsmuth, Syed, and
  Stein}]{wachsmuth2018retrieval}
Henning Wachsmuth, Shahbaz Syed, and Benno Stein. 2018.
\newblock \href {https://doi.org/10.18653/v1/P18-1023} {Retrieval of the best
  counterargument without prior topic knowledge}.
\newblock In \emph{Proceedings of the 56th Annual Meeting of the Association
  for Computational Linguistics (Volume 1: Long Papers)}, pages 241--251,
  Melbourne, Australia. Association for Computational Linguistics.

\bibitem[{Wadden et~al.(2020)Wadden, Lin, Lo, Wang, van Zuylen, Cohan, and
  Hajishirzi}]{wadden2020fact}
David Wadden, Shanchuan Lin, Kyle Lo, Lucy~Lu Wang, Madeleine van Zuylen, Arman
  Cohan, and Hannaneh Hajishirzi. 2020.
\newblock \href {https://doi.org/10.18653/v1/2020.emnlp-main.609} {Fact or
  fiction: Verifying scientific claims}.
\newblock In \emph{Proceedings of the 2020 Conference on Empirical Methods in
  Natural Language Processing (EMNLP)}, pages 7534--7550, Online. Association
  for Computational Linguistics.

\bibitem[{Wang et~al.(2020)Wang, Wei, Dong, Bao, Yang, and
  Zhou}]{wang2020minilm}
Wenhui Wang, Furu Wei, Li~Dong, Hangbo Bao, Nan Yang, and Ming Zhou. 2020.
\newblock \href {https://arxiv.org/abs/2002.10957} {{M}ini{LM}: {D}eep
  {S}elf-{A}ttention {D}istillation for {T}ask-{A}gnostic {C}ompression of
  {P}re-{T}rained {T}ransformers}.
\newblock \emph{arXiv preprint arXiv:2002.10957}.

\bibitem[{Wei et~al.(2014)Wei, Guan, and Yu}]{wei2014projected}
Benchang Wei, Tao Guan, and Junqing Yu. 2014.
\newblock {P}rojected {R}esidual {V}ector {Q}uantization for {ANN} {S}earch.
\newblock \emph{IEEE multimedia}, 21(3):41--51.

\bibitem[{Wolf et~al.(2020)Wolf, Debut, Sanh, Chaumond, Delangue, Moi, Cistac,
  Rault, Louf, Funtowicz, Davison, Shleifer, von Platen, Ma, Jernite, Plu, Xu,
  Le~Scao, Gugger, Drame, Lhoest, and Rush}]{wolf2020transformers}
Thomas Wolf, Lysandre Debut, Victor Sanh, Julien Chaumond, Clement Delangue,
  Anthony Moi, Pierric Cistac, Tim Rault, Remi Louf, Morgan Funtowicz, Joe
  Davison, Sam Shleifer, Patrick von Platen, Clara Ma, Yacine Jernite, Julien
  Plu, Canwen Xu, Teven Le~Scao, Sylvain Gugger, Mariama Drame, Quentin Lhoest,
  and Alexander Rush. 2020.
\newblock \href {https://doi.org/10.18653/v1/2020.emnlp-demos.6} {Transformers:
  State-of-the-art natural language processing}.
\newblock In \emph{Proceedings of the 2020 Conference on Empirical Methods in
  Natural Language Processing: System Demonstrations}, pages 38--45, Online.
  Association for Computational Linguistics.

\bibitem[{Xin et~al.(2021)Xin, Xiong, Srinivasan, Sharma, Jose, and
  Bennett}]{xin2021zero}
Ji~Xin, Chenyan Xiong, Ashwin Srinivasan, Ankita Sharma, Damien Jose, and
  Paul~N Bennett. 2021.
\newblock \href {https://arxiv.org/abs/2110.07581} {{Z}ero-{S}hot {D}ense
  {R}etrieval with {M}omentum {A}dversarial {D}omain {I}nvariant
  {R}epresentations}.
\newblock \emph{arXiv preprint arXiv:2110.07581}.

\bibitem[{Xiong et~al.(2020)Xiong, Xiong, Li, Tang, Liu, Bennett, Ahmed, and
  Overwijk}]{xiong2020approximate}
Lee Xiong, Chenyan Xiong, Ye~Li, Kwok-Fung Tang, Jialin Liu, Paul~N Bennett,
  Junaid Ahmed, and Arnold Overwijk. 2020.
\newblock {A}pproximate {N}earest {N}eighbor {N}egative {C}ontrastive
  {L}earning for {D}ense {T}ext {R}etrieval.
\newblock In \emph{International Conference on Learning Representations}.

\bibitem[{Yamada et~al.(2021{\natexlab{a}})Yamada, Asai, and
  Hajishirzi}]{yamada2021efficient}
Ikuya Yamada, Akari Asai, and Hannaneh Hajishirzi. 2021{\natexlab{a}}.
\newblock \href {https://doi.org/10.18653/v1/2021.acl-short.123} {Efficient
  passage retrieval with hashing for open-domain question answering}.
\newblock In \emph{Proceedings of the 59th Annual Meeting of the Association
  for Computational Linguistics and the 11th International Joint Conference on
  Natural Language Processing (Volume 2: Short Papers)}, pages 979--986,
  Online. Association for Computational Linguistics.

\bibitem[{Yamada et~al.(2021{\natexlab{b}})Yamada, Asai, and
  Hajishirzi}]{yamada-etal-2021-efficient}
Ikuya Yamada, Akari Asai, and Hannaneh Hajishirzi. 2021{\natexlab{b}}.
\newblock \href {https://doi.org/10.18653/v1/2021.acl-short.123} {Efficient
  passage retrieval with hashing for open-domain question answering}.
\newblock In \emph{Proceedings of the 59th Annual Meeting of the Association
  for Computational Linguistics and the 11th International Joint Conference on
  Natural Language Processing (Volume 2: Short Papers)}, pages 979--986,
  Online. Association for Computational Linguistics.

\bibitem[{Yang et~al.(2018{\natexlab{a}})Yang, Fang, and
  Lin}]{yang2018anserini}
Peilin Yang, Hui Fang, and Jimmy Lin. 2018{\natexlab{a}}.
\newblock Anserini: Reproducible ranking baselines using lucene.
\newblock \emph{Journal of Data and Information Quality (JDIQ)}, 10(4):1--20.

\bibitem[{Yang et~al.(2018{\natexlab{b}})Yang, Qi, Zhang, Bengio, Cohen,
  Salakhutdinov, and Manning}]{yang-etal-2018-hotpotqa}
Zhilin Yang, Peng Qi, Saizheng Zhang, Yoshua Bengio, William Cohen, Ruslan
  Salakhutdinov, and Christopher~D. Manning. 2018{\natexlab{b}}.
\newblock \href {https://doi.org/10.18653/v1/D18-1259} {{H}otpot{QA}: A dataset
  for diverse, explainable multi-hop question answering}.
\newblock In \emph{Proceedings of the 2018 Conference on Empirical Methods in
  Natural Language Processing}, pages 2369--2380, Brussels, Belgium.
  Association for Computational Linguistics.

\bibitem[{Zhan et~al.(2021{\natexlab{a}})Zhan, Mao, Liu, Guo, Zhang, and
  Ma}]{zhan2021jointly}
Jingtao Zhan, Jiaxin Mao, Yiqun Liu, Jiafeng Guo, Min Zhang, and Shaoping Ma.
  2021{\natexlab{a}}.
\newblock {J}ointly {O}ptimizing {Q}uery {E}ncoder and {P}roduct {Q}uantization
  to {I}mprove {R}etrieval {P}erformance.
\newblock In \emph{Proceedings of the 30th ACM International Conference on
  Information \& Knowledge Management}, pages 2487--2496.

\bibitem[{Zhan et~al.(2021{\natexlab{b}})Zhan, Mao, Liu, Guo, Zhang, and
  Ma}]{zhan2021optimizing}
Jingtao Zhan, Jiaxin Mao, Yiqun Liu, Jiafeng Guo, Min Zhang, and Shaoping Ma.
  2021{\natexlab{b}}.
\newblock {O}ptimizing {D}ense {R}etrieval {M}odel {T}raining with {H}ard
  {N}egatives.
\newblock In \emph{Proceedings of the 44th International ACM SIGIR Conference
  on Research and Development in Information Retrieval}, pages 1503--1512.

\bibitem[{Zhan et~al.(2022)Zhan, Mao, Liu, Guo, Zhang, and
  Ma}]{zhan2021learning}
Jingtao Zhan, Jiaxin Mao, Yiqun Liu, Jiafeng Guo, Min Zhang, and Shaoping Ma.
  2022.
\newblock \href {https://doi.org/10.1145/3488560.3498443} {Learning discrete
  representations via constrained clustering for effective and efficient dense
  retrieval}.
\newblock In \emph{Proceedings of the Fifteenth ACM International Conference on
  Web Search and Data Mining}, WSDM '22, page 1328–1336. Association for
  Computing Machinery.

\bibitem[{Zhan et~al.(2020{\natexlab{a}})Zhan, Mao, Liu, Zhang, and
  Ma}]{zhan2020learning}
Jingtao Zhan, Jiaxin Mao, Yiqun Liu, Min Zhang, and Shaoping Ma.
  2020{\natexlab{a}}.
\newblock \href {https://arxiv.org/abs/2010.10469} {Learning to retrieve: How
  to train a dense retrieval model effectively and efficiently}.
\newblock \emph{arXiv preprint arXiv:2010.10469}.

\bibitem[{Zhan et~al.(2020{\natexlab{b}})Zhan, Mao, Liu, Zhang, and
  Ma}]{zhan2020repbert}
Jingtao Zhan, Jiaxin Mao, Yiqun Liu, Min Zhang, and Shaoping Ma.
  2020{\natexlab{b}}.
\newblock \href {https://arxiv.org/abs/2006.15498} {Repbert: Contextualized
  text embeddings for first-stage retrieval}.
\newblock \emph{arXiv preprint arXiv:2006.15498}.

\bibitem[{Zhou and Devlin(2021)}]{zhou-devlin-2021-multi}
Giulio Zhou and Jacob Devlin. 2021.
\newblock \href {https://doi.org/10.18653/v1/2021.emnlp-main.443} {Multi-vector
  attention models for deep re-ranking}.
\newblock In \emph{Proceedings of the 2021 Conference on Empirical Methods in
  Natural Language Processing}, pages 5452--5456, Online and Punta Cana,
  Dominican Republic. Association for Computational Linguistics.

\end{thebibliography}
\bibliographystyle{acl_natbib}

\appendix

\section{Analysis of ColBERT's Semantic Space} \label{sec:analysis}

ColBERT~\cite{khattab2020colbert} decomposes representations and similarity computation at the token level. Because of this compositional architecture, we hypothesize that ColBERT exhibits a ``lightweight'' semantic space: without any special re-training, vectors corresponding to each sense of a word would cluster very closely, with only minor variation due to context.

\begin{figure}[tp]
\centering
\begin{subfigure}[b]{0.65\columnwidth}
\centering
\includegraphics[width=\columnwidth]{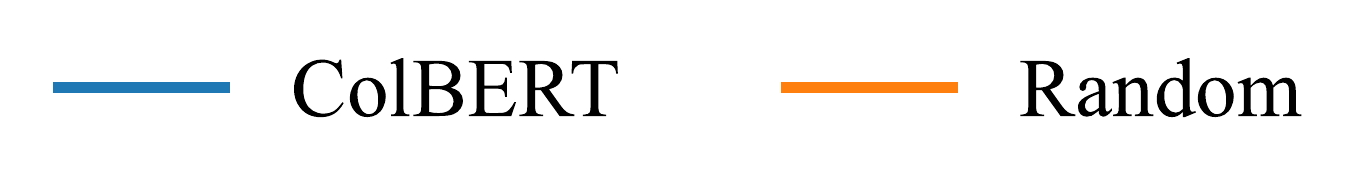}
\end{subfigure}
\begin{subfigure}[b]{0.48\columnwidth}
\centering
\includegraphics[width=\columnwidth]{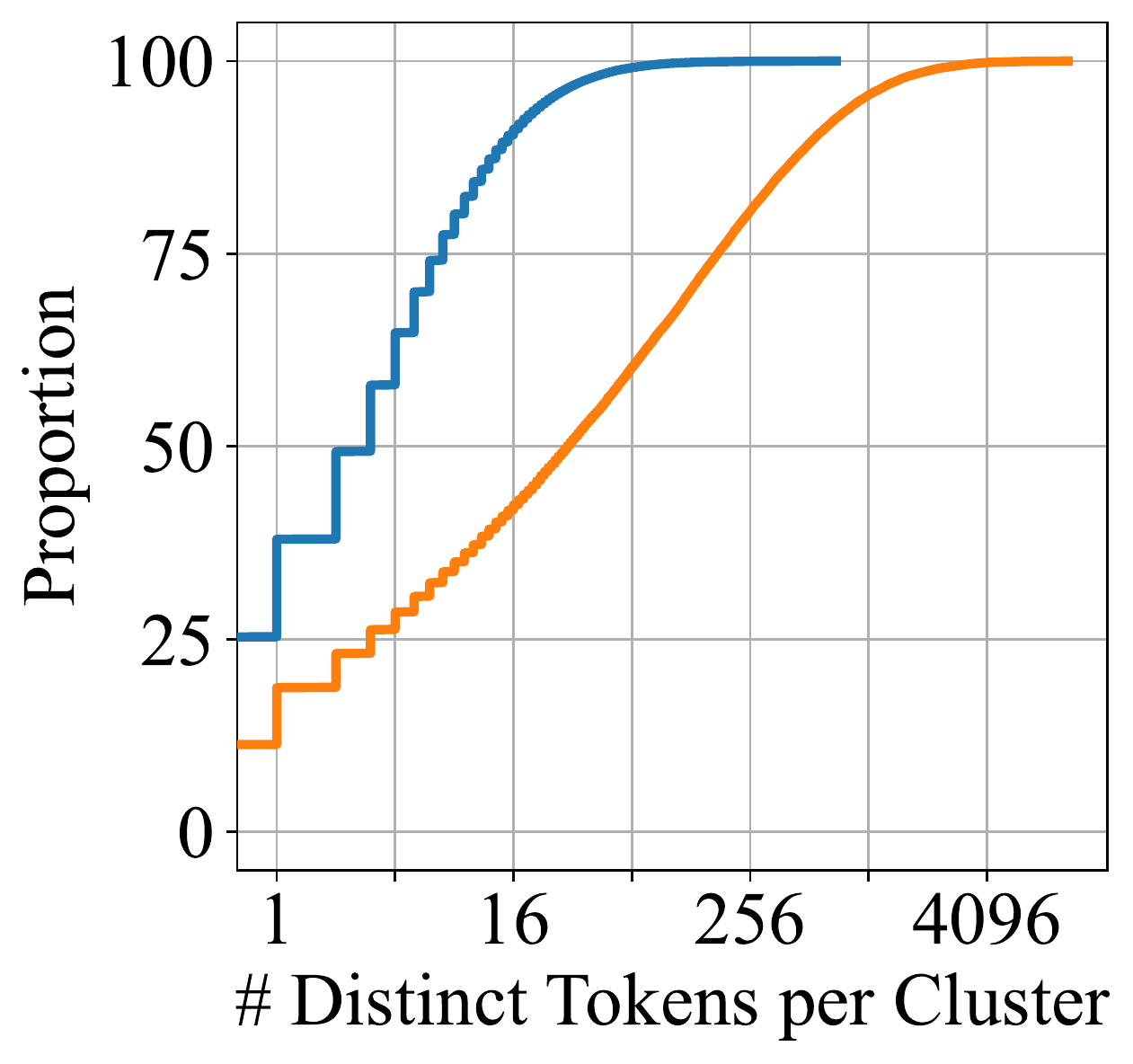}
\caption{Number of distinct tokens appearing in each cluster.}
\label{fig:tokens_per_cluster}
\end{subfigure}
\begin{subfigure}[b]{0.48\columnwidth}
\centering
\includegraphics[width=\columnwidth]{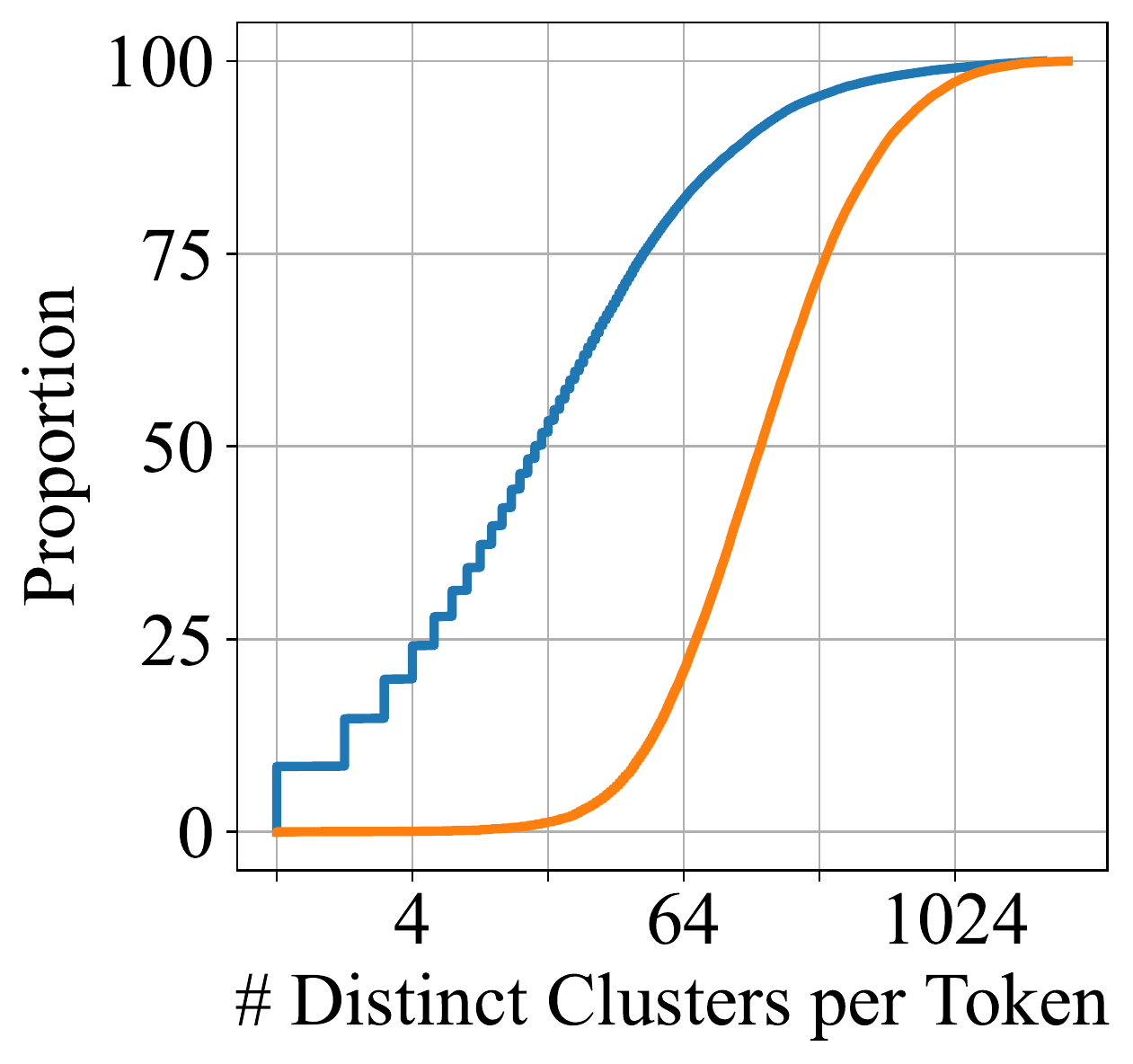}
\caption{Number of distinct clusters each token appears in.}
\label{fig:clusters_per_token}
\end{subfigure}
\caption{Empirical CDFs analyzing semantic properties of MS MARCO token-level embeddings both encoded by ColBERT and randomly generated. The embeddings are partitioned into $2^{18}$ clusters and correspond to roughly 27,000 distinct tokens.}
\label{fig:semantic_space_analysis}
\end{figure}

\begin{table*}[]
\centering
\resizebox{\linewidth}{!}{
\begin{tabular}{@{}rccc@{}}
\toprule
\multicolumn{1}{c}{\multirow{2}{*}{Cluster ID}} & \multirow{2}{*}{Most Common Tokens}                                                                                            & \multicolumn{2}{c}{Most Common Clusters Per Token}                                                                                 \\ \cmidrule(l){3-4} 
\multicolumn{1}{c}{}                            &                                                                                                                        & Token      & Clusters                                                                                                      \\ \midrule
\multirow{3}{*}{917}                           & \multirow{3}{*}{\begin{tabular}[c]{@{}c@{}}`photos', `photo', `pictures',\\ `photographs', `images',\\`photography', `photograph'\end{tabular}} & `photos'   & \begin{tabular}[c]{@{}c@{}}Photos-Photo, Photos-Pictures-Photo\end{tabular}   \\ \cmidrule(l){3-4}                                 
                                                &                                                                                                                        & `photo'    & \begin{tabular}[c]{@{}c@{}}Photo-Image-Picture, Photo-Picture-Photograph, Photo-Picture-Photography\end{tabular}                              \\ \cmidrule(l){3-4}
                                                &                                                                                                                        & `pictures'  & \begin{tabular}[c]{@{}c@{}}Pictures-Picture-Images, Picture-Pictures-Artists, Pictures-Photo-Picture\end{tabular} \\ \midrule 
\multirow{3}{*}{216932}                          & \multirow{3}{*}{\begin{tabular}[c]{@{}c@{}}`tornado', `tornadoes', `storm'\\ `hurricane', `storms'\end{tabular}}                       & `tornado'   & \begin{tabular}[c]{@{}c@{}} Tornado-Hurricane-Storm, Tornadoes-Tornado-Blizzard\end{tabular}                                       \\ \cmidrule(l){3-4}
                                                &                                                                                                                        & `tornadoes' & \begin{tabular}[c]{@{}c@{}}Tornadoes-Tornado-Storms, Tornadoes-Tornado-Blizzard, Tornado-Hurricane-Storm\end{tabular}                                     \\ \cmidrule(l){3-4}
                                                &                                                                                                                        & `storm'    & \begin{tabular}[c]{@{}c@{}}Storm-Storms, Storm-Storms-Weather, Storm-Storms-Tempest\end{tabular}                       \\ \bottomrule
\end{tabular}
}
\caption{Examples of clusters taken from all MS MARCO passages. We present the tokens that appear most frequently in the selected clusters as well as additional clusters the top tokens appear in.}
\label{table:semantic_space_analysis}
\end{table*}

If this hypothesis is true, we would expect the embeddings corresponding to each token in the vocabulary to localize in only a small number of regions in the embedding space, corresponding to the contextual ``senses'' of the token. To validate this hypothesis, we analyze the ColBERT embeddings corresponding to the tokens in the MS MARCO Passage Ranking~\cite{nguyen2016ms} collection: we perform $k$-means clustering on the nearly 600M embeddings---corresponding to 27,000 unique tokens---into $k=2^{18}$ clusters. As a baseline, we repeat this clustering with random embeddings but keep the true distribution of tokens. Figure~\ref{fig:semantic_space_analysis} presents empirical cumulative distribution function (eCDF) plots representing the number of distinct non-stopword tokens appearing in each cluster (\ref{fig:tokens_per_cluster}) and the number of distinct clusters in which each token appears (\ref{fig:clusters_per_token}).\footnote{We rank tokens by number of clusters they appear in and designate the top-1\% (under 300) as stopwords.} Most tokens appear in a very small fraction of the number of centroids: in particular, we see that roughly 90\% of clusters have $\le$ 16 distinct tokens with the ColBERT embeddings, whereas less than 50\% of clusters have $\le$ 16 distinct tokens with the random embeddings. This suggests that the centroids effectively map the ColBERT semantic space. 

Table~\ref{table:semantic_space_analysis} presents examples to highlight the semantic space captured by the centroids. The most frequently appearing tokens in cluster \#917 relate to photography; these include, for example, `photos' and `photographs'. If we then examine the additional clusters in which these tokens appear, we find that there is substantial semantic overlap between these new clusters (e.g., Photos-Photo, Photo-Image-Picture) and cluster \#917. We observe a similar effect with tokens appearing in cluster \#216932, comprising tornado-related terms.

This analysis indicates that cluster centroids can summarize the ColBERT representations with high precision. In \secref{sec:representation}, we propose a residual compression mechanism that uses these centroids along with minor refinements at the dimension level to efficiently encode late-interaction vectors. %

\section{Impact of Compression}
\label{appendix:compression}

Our residual compression approach (\secref{sec:representation}) preserves approximately the same quality as the uncompressed embeddings. In particular, when applied to a vanilla ColBERT model on MS MARCO whose MRR@10 is 36.2\% and Recall@50 is 82.1\%, the quality of the model with 2-bit compression is 36.2\% MRR@10 and 82.3\% Recall@50. With 1-bit compression, the model achieves 35.5\% MRR@10 and 81.6\% Recall@50.\footnote{We contrast this with an early implementation of compression for ColBERT, which used binary representations as in BPR~\cite{yamada2021efficient} without residual centroids, and achieves 34.8\% (35.7\%) MRR@10 and 80.5\% (81.8\%) Recall@50 with 1-bit (2-bit) binarization. Like the original ColBERT, this form of compression relied on a separate FAISS index for candidate generation.}

\begin{figure*}[]
\centering
\includegraphics[width=\textwidth]{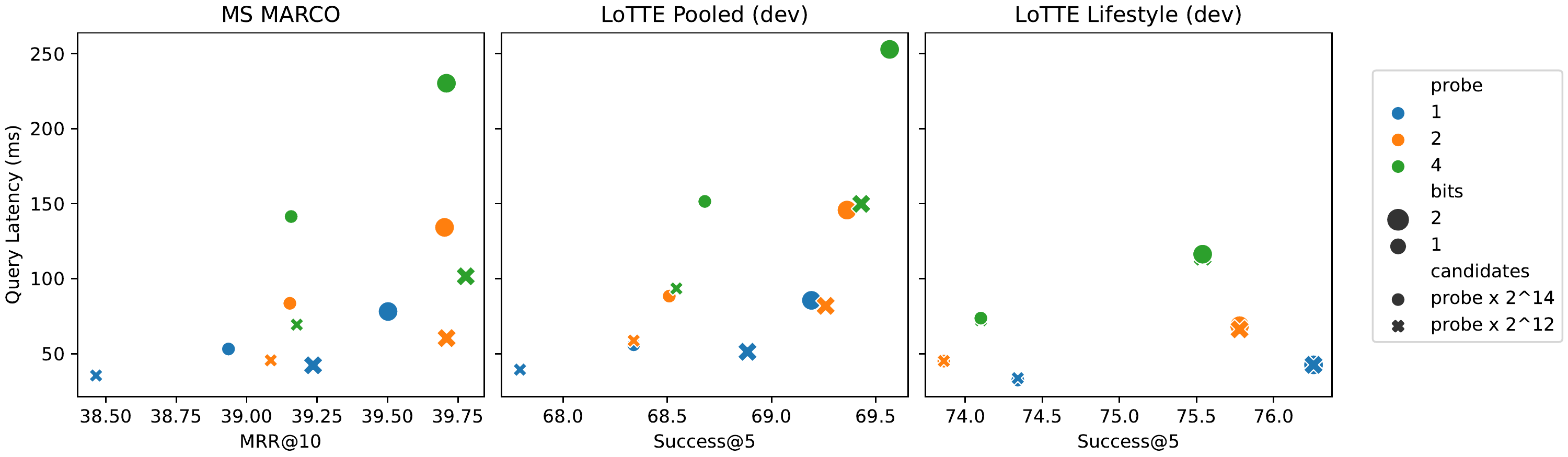}
\caption{Latency vs. retrieval quality with varying parameter configurations for three datasets of different collection sizes. We sweep a range of values for the number of centroids per vector ($\texttt{probe}$), the number of bits used for residual compression, and the number of candidates. Note that retrieval quality is measured in MRR@10 for MS MARCO and Success@5 for \dataset{} datasets. Results toward the bottom right corner (higher quality, lower latency) are best.}
\label{fig:latency}
\end{figure*}

We also tested the residual compression approach on late-interaction retrievers that conduct downstream tasks, namely, ColBERT-QA~\cite{khattab2021relevance} for the NaturalQuestions open-domain QA task, and Baleen~\cite{khattab2021baleen} for multi-hop reasoning on HoVer for claim verification. On the NQ dev set, ColBERT-QA's success@5 (success@20) dropped only marginally from 75.3\% (84.3\%) to 74.3\% (84.2\%) and its downstream Open-QA answer exact match dropped from 47.9\% to 47.7\%, when using 2-bit compression for retrieval and using the same checkpoints of ColBERT-QA otherwise.

Similarly, on the HoVer~\cite{jiang2020hover} dev set, Baleen's retrieval R@100 dropped from 92.2\% to only 90.6\% but its sentence-level exact match remained roughly the same, going from 39.2\% to 39.4\%. We hypothesize that the supervision methods applied in \system{} (\secref{sec:supervision}) can also be applied to lift quality in downstream tasks by improving the recall of retrieval for these tasks. We leave such exploration for future work.

\section{Retrieval Latency}
\label{appendix:latency}

Figure~\ref{fig:latency} evaluates the latency of \system{} across three collections of varying sizes, namely, MS MARCO, \dataset{} Pooled (dev), and \dataset{} Lifestyle (dev), which contain approximately 9M passages, 2.4M answer posts, and 270k answer posts, respectively. We average latency across three runs of the MS MARCO dev set and the \dataset{} ``search'' queries. Search is executed using a Titan V GPU on a server with two Intel Xeon Gold 6132 CPUs, each with 28 hardware execution contexts.

The figure varies three settings of \system{}. In particular, we evaluate indexing with 1-bit and 2-bit encoding (\secref{sec:indexing}) and searching by probing the nearest 1, 2, or 4 centroids to each query vector (\secref{sec:retrieval}). When probing \texttt{probe} centroids per vector, we score either $\texttt{probe} \times 2^{12}$ or $\texttt{probe} \times 2^{14}$ candidates per query.\footnote{These settings are selected based on preliminary exploration of these parameters, which indicated that performance for larger \texttt{probe} values tends to require scoring a larger number of candidates.}

To begin with, we notice that the quality reported on the $x$-axis varies only within a relatively narrow range. For instance, the axis ranges from 38.50 through 39.75 for MS MARCO, and all but two of the cheapest settings score above 39.00. Similarly, the $y$-axis varies between approximately 50 milliseconds per query up to 250 milliseconds (mostly under 150 milliseconds) using our relatively simple Python-based implementation.

Digging deeper, we see that the best quality in these metrics can be achieved or approached closely with around 100 milliseconds of latency across all three datasets, despite their various sizes and characteristics, and that 2-bit indexing reliably outperforms 1-bit indexing but the loss from more aggressive compression is small.

\section{\dataset{}} \label{section:appendix_lotte}

\begin{figure}[t]
    \centering
    \includegraphics[width=\columnwidth]{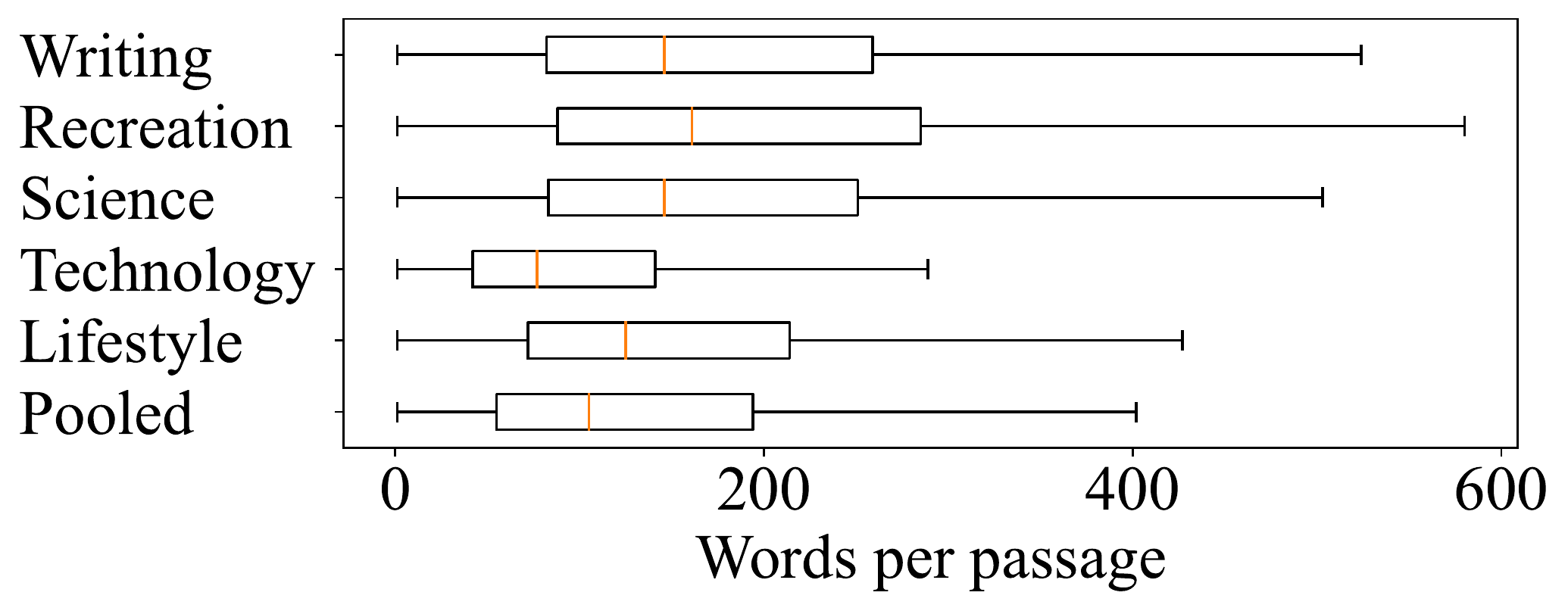}
    \caption{\dataset{} words per passage}
    \label{fig:lotte_words_per_passage}
\end{figure}

\begin{figure}[t]
    \centering
    \includegraphics[width=\columnwidth]{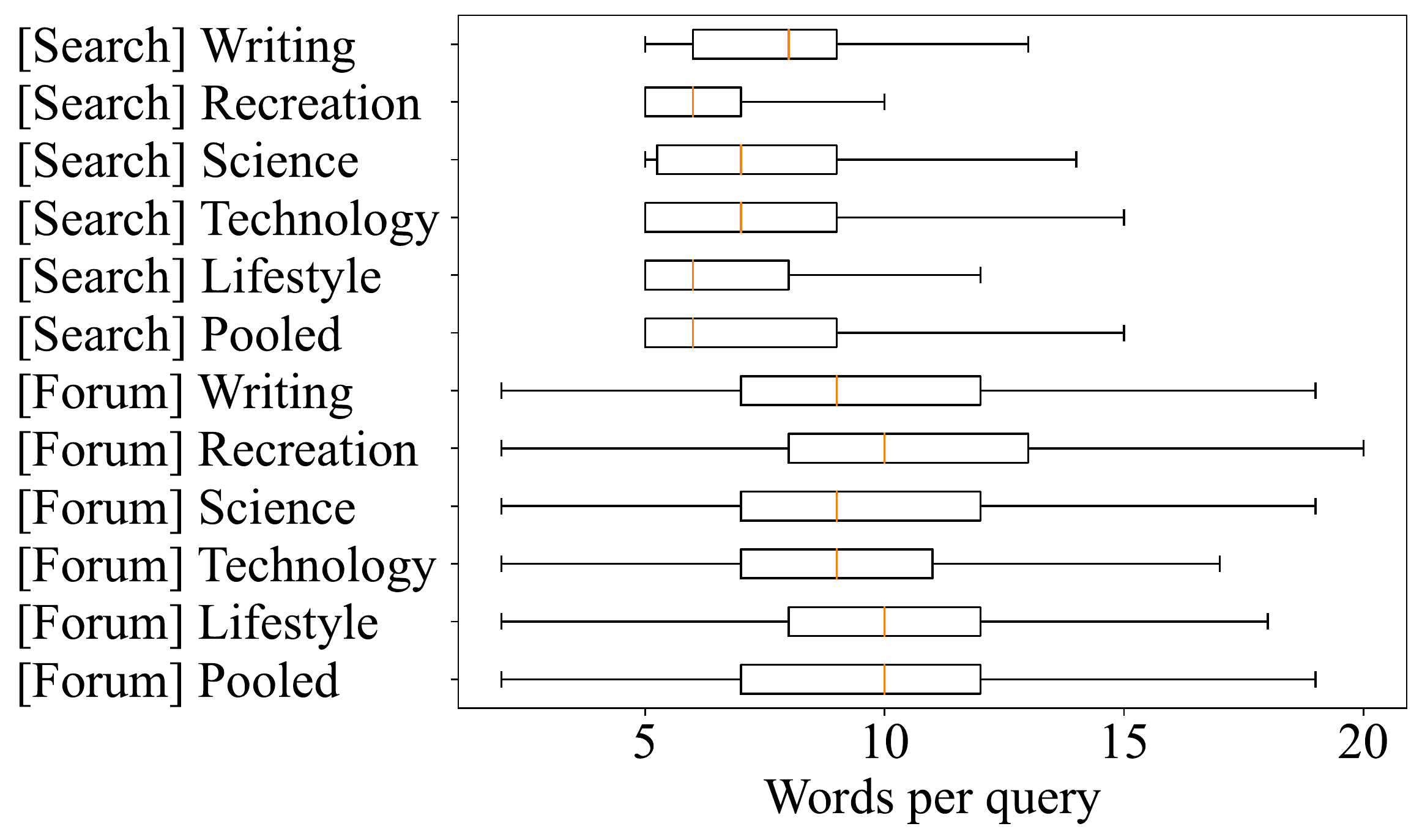}
    \caption{\dataset{} words per query}
    \label{fig:lotte_words_per_query}
\end{figure}

\begin{figure}[t]
    \centering
    \includegraphics[width=\columnwidth]{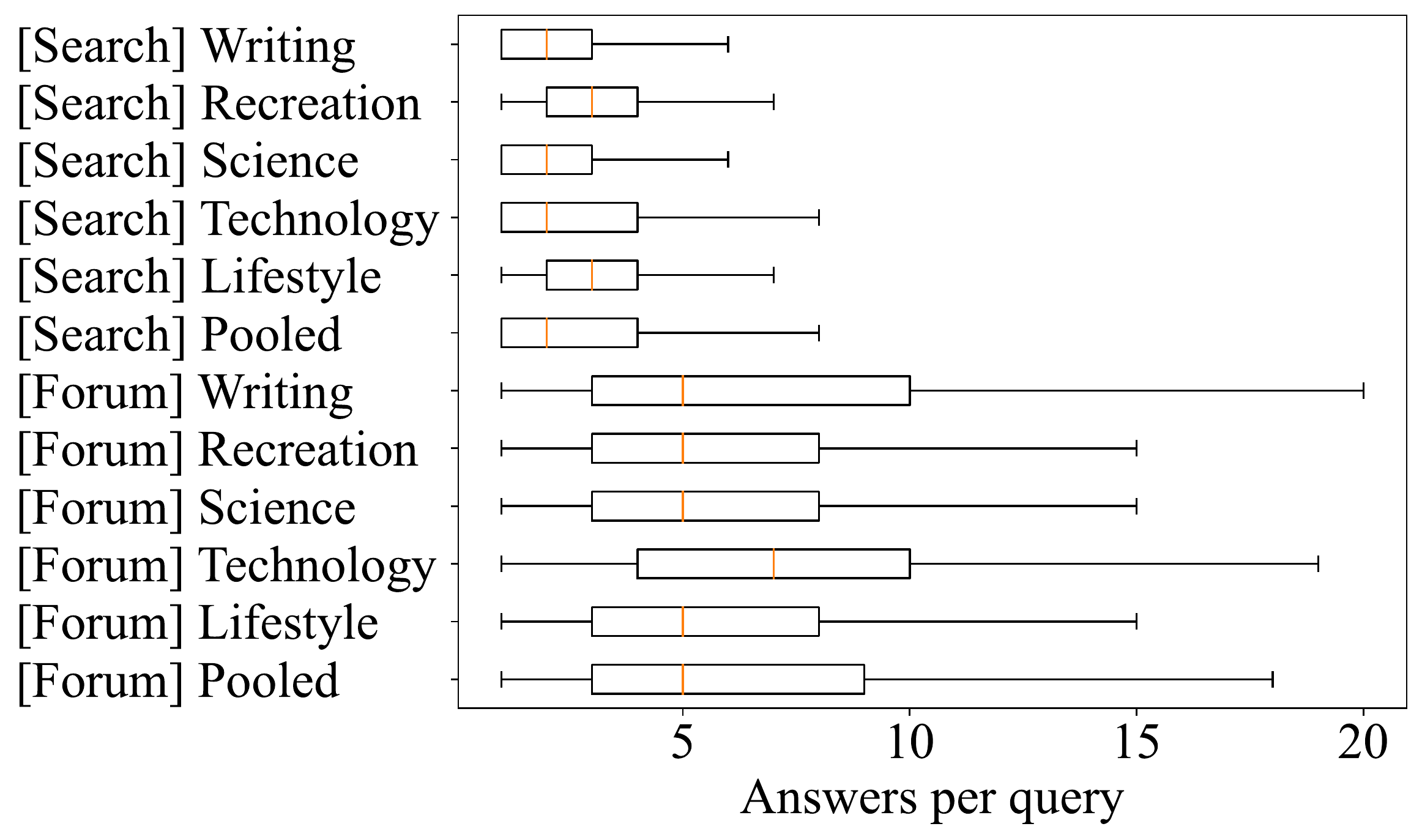}
    \caption{\dataset{} answers per query}
    \label{fig:lotte_answers_per_query}
\end{figure}

\paragraph{Domain coverage} Table~\ref{table:lotte_communities} presents the full distribution of communities in the \dataset{} dev dataset. The topics covered by \dataset{} cover a wide range of linguistic phenomena given the diversity in topics and communities represented. However, since all posts are submitted by anonymous users we do not have demographic information regarding the identify of the contributors. All posts are written in English.

\paragraph{Passages} As mentioned in \S\ref{section:sedata}, we construct \dataset{} collections by selecting passages from the StackExchange archive with positive scores. We remove HTML tags from passages and filter out empty passages. For each passage we record its corresponding query and save the query-to-passage mapping to keep track of the posted answers corresponding to each query.

\paragraph{Search queries} We construct the list of \dataset{} search queries by drawing from GooAQ queries that appear in the StackExchange post archive. We first shuffle the list of GooAQ queries so that in cases where multiple queries exist for the same answer passage we randomly select the query to include in \dataset{} rather than always selecting the first appearing query. We verify that every query has at least one corresponding answer passage.

\paragraph{Forum queries} For each \dataset{} topic and its constituent communities we first compute the fraction of the total queries attributed to each individual community. We then use this distribution to construct a truncated query set by selecting the highest ranked queries from each community as determined by 1) the query scores and 2) the query view counts. We only use queries which have an accepted answer. We ensure that each community contributes at least 50 queries to the truncated set whenever possible. We set the overall size of the truncated set to be 2000 queries, though note that the total can exceed this due to rounding and/or the minimum per-community query count. We remove all quotation marks and HTML tags.

\paragraph{Statistics} Figure~\ref{fig:lotte_words_per_passage} plots the number of words per passage in each \dataset{} dev corpus. Figures~\ref{fig:lotte_words_per_query} and~\ref{fig:lotte_answers_per_query} plot the number of words and number of corresponding answer passages respectively per query, split across search and forum queries. 

\begin{table}
\centering
\footnotesize
\setlength{\tabcolsep}{4pt}
\begin{tabular}[b]{@{}lcccccc@{}}
\toprule
\multicolumn{1}{c}{Corpus} & \multicolumn{1}{c}{\rotatebox[origin=c]{270}{\scalebox{.85}{\textbf{ColBERT}}}} &
\multicolumn{1}{c}{\rotatebox[origin=c]{270}{\scalebox{.85}{\textbf{BM25}}}} &
\multicolumn{1}{c}{\rotatebox[origin=c]{270}{\scalebox{.85}{\textbf{ANCE}}}} &
\multicolumn{1}{c}{\rotatebox[origin=c]{270}{\scalebox{.85}{\textbf{RocketQAv2}}}} &
\multicolumn{1}{c}{\rotatebox[origin=c]{270}{\scalebox{.85}{\textbf{SPLADEv2}}}} & \multicolumn{1}{c}{\rotatebox[origin=c]{270}{\scalebox{.85}{\textbf{\system{}}}}} \\ \midrule 
                           \multicolumn{5}{c}{\dataset{} Search Dev Queries (Success@5)}                                        \\ \midrule
\scalebox{.85}{\textbf{Writing}}                    & 76.3                  & 47.3                                & 75.7                    & 79.5                      & 78.9                         & \textbf{81.7}             \\
\scalebox{.85}{\textbf{Recreation}}              & 71.8                           & 56.3                          & 66.1                    & 73.0                      & 70.7                         & \textbf{76.0}             \\
\scalebox{.85}{\textbf{Science}}                  & 71.7                        & 52.2                            & 66.9                    & 67.7                      & 73.4                         & \textbf{74.2}             \\
\scalebox{.85}{\textbf{Technology}}                 & 52.8                      & 35.8                            & 55.7                    & 54.3                      & 56.3                         & \textbf{59.3}             \\
\scalebox{.85}{\textbf{Lifestyle}}                  & 73.1                  & 54.4                                & 69.8                    & 72.4                      & 71.2                     & \textbf{75.8}             \\
\scalebox{.85}{\textbf{Pooled}}                     & 65.4                  & 45.6                                & 63.7                    & 66.4                   & 67.0                         & \textbf{69.3}             \\ \midrule
                           \multicolumn{5}{c}{\dataset{} Forum Dev Queries (Success@5)}                                         \\ \midrule
\scalebox{.85}{\textbf{Writing}}                    & 75.5                        & 66.2                          & 74.4                    & 75.5                      & 78.1                         & \textbf{80.8}             \\
\scalebox{.85}{\textbf{Recreation}}              & 69.1                         & 56.6                         & 65.9                       & 69.0                      & 68.9                         & \textbf{71.8}             \\
\scalebox{.85}{\textbf{Science}}                  & 58.2                        & 51.3                          & 56.3                  & 56.7                      & 59.9                         & \textbf{62.6}             \\
\scalebox{.85}{\textbf{Technology}}                 & 39.6                      & 30.7                           & 38.8                    & 39.9                     & 42.1                         & \textbf{45.0}             \\
\scalebox{.85}{\textbf{Lifestyle}}                  & 61.1                  & 48.2                               & 61.8                    & 62.0                      & 61.8                         & \textbf{65.8}             \\
\scalebox{.85}{\textbf{Pooled}}                     & 59.1                  & 47.8                                & 57.4                    & 58.9                    & 60.6                         & \textbf{63.7}             \\ \bottomrule
\end{tabular}
\vspace{-3mm}
\caption{Zero-shot evaluation results on the dev sets of the \dataset{} benchmark.}
\label{table:lotte_dev_results}
\vspace{-3mm}
\end{table}
\paragraph{Dev Results} Table~\ref{table:lotte_dev_results} presents out-of-domain evaluation results on the \dataset{} dev queries. Continuing the trend we observed in~\ref{sec:evaluation}, \system{} consistently outperforms all other models we tested.

\paragraph{Licensing and Anonymity} The original StackExchange post archive is licensed under a Creative Commons BY-SA 4.0 license~\cite{stackexchange}. Personal data is removed from the archive before being uploaded, though all posts are public; when we release \dataset{} publicly we will include URLs to the original posts for proper attribution as required by the license. The GooAQ dataset is licensed under an Apache license, version 2.0~\cite{khashabi2021gooaq}. We will also release \dataset{} with a CC BY-SA 4.0 license. The search queries can be used for non-commercial research purposes only as per the GooAQ license.

\section{Datasets in BEIR}
\label{appendix:beir}

Table~\ref{table:beir_dataset_info} lists the BEIR datasets we used in our evaluation, including their respective license information as well as the numbers of documents as well as the number of test set queries. We refer to \citet{thakur2021beir} for a more detailed description of each of the datasets.

Our Touch\'e evaluation uses an updated version of the data in BEIR, which we use for evaluating the models we run (i.e., \system{} and RocketQAv2) as well as SPLADEv2.

\begin{table}[h]
\resizebox{\columnwidth}{!}{%
\begin{tabular}{@{}lrrrrr@{}}
\toprule
\multicolumn{1}{c}{Dataset} & \multicolumn{1}{c}{License}                                               & \multicolumn{1}{c}{\# Passages}  & \multicolumn{1}{c}{\# Test Queries} \\ \midrule
ArguAna~\cite{wachsmuth2018retrieval}                                                                   & CC BY 4.0                                                                 & 8674                             & 1406                     \\
Climate-Fever~\cite{diggelmann2020climate}                                                              & Not reported                                                              & 5416593                          & 1535                     \\
DBPedia~\cite{auer2007dbpedia}                                                                          & CC BY-SA 3.0                                                              & 4635922                          & 400                      \\
FEVER~\cite{thorne2018fever}                                                                            & CC BY-SA 3.0                                                              &                                  &                          \\
FiQA-2018~\cite{maia2018fiqa}                                                                           & Not reported                                                              & 57638                            & 648                      \\
HotpotQA~\cite{yang-etal-2018-hotpotqa}                                                                 & CC BY-SA 4.0                                                              & 5233329                          & 7405                     \\
NFCorpus~\cite{boteva2016full}                                                                          & Not reported                                                              & 3633                             & 323                      \\
NQ~\cite{kwiatkowski2019natural}                                                                        & CC BY-SA 3.0                                                              & 2681468                          & 3452                     \\
SCIDOCS~\cite{cohan2020specter}                                                                         & \begin{tabular}[c]{@{}r@{}}GNU General Public\\ License v3.0\end{tabular} & 25657                            & 1000                     \\
SciFact~\cite{wadden2020fact}                                                                           & CC BY-NC 2.0                                                              & 5183                             & 300                      \\
Quora                                                                                                   & Not reported                                                              & 522931                           & 10000                    \\
Touch\'e-2020~\cite{bondarenko2020overview}                                                             & CC BY 4.0                                                                 & 382545                           & 49                       \\
TREC-COVID~\cite{voorhees2021trec}                                                                      & \begin{tabular}[c]{@{}r@{}}Dataset License\\ Agreement\end{tabular}       & 171332                           & 50                       \\
\bottomrule
\end{tabular}
}
\caption{BEIR dataset information.}
\label{table:beir_dataset_info}
\end{table}

We also tested on the Open-QA benchmarks NQ, TQ, and SQuAD, each of which has approximately 9k dev-set questions and muli-hop HoVer, whose development set has 4k claims. In the compression evaluation~\secref{appendix:compression}, we used models trained in-domain on NQ and HoVer, whose training sets contain 79k and 18k queries, respectively.

\section{Implementation \& Hyperparameters}

We implement \system{} using Python 3.7, PyTorch 1.9, and HuggingFace Transformers 4.10~\cite{wolf2020transformers}, extending the original implementation of ColBERT by \citet{khattab2020colbert}. We use FAISS 1.7~\cite{johnson2019billion} for $k$-means clustering,\footnote{\url{https://github.com/facebookresearch/faiss}} though unlike ColBERT we do not use it for nearest-neighbor search. Instead, we implement our candidate generation mechanism (\secref{sec:retrieval}) using PyTorch primitives in Python.

We conducted our experiments on an internal cluster, typically using up to four 12GB Titan V GPUs for each of the inference tasks (e.g., indexing, computing distillation scores, and retrieval) and four 80GB A100 GPUs for training, though GPUs with smaller RAM can be used via gradient accumulation. Using this infrastructure, computing the distillation scores takes under a day, training a 64-way model on MS MARCO for 400,000 steps takes around five days, and indexing takes approximately two hours. We very roughly estimate an upper bound total of 20 GPU-months for all experimentation, development, and evaluation performed for this work over a period of several months.

Like ColBERT, our encoder is a \texttt{bert-base-uncased} model that is shared between the query and passage encoders and which has 110M parameters. We retain the default vector dimension suggested by \citet{khattab2020colbert} and used in subsequent work, namely, $d$=128. For the experiments reported in this paper, we train on MS MARCO training set. We use simple defaults with limited manual exploration on the official development set for the learning rate ($10^{-5}$), batch size (32 examples), and warm up (for 20,000 steps) with linear decay.

Hyperparameters corresponding to retrieval are explored in~\secref{appendix:latency}. We default to $\texttt{probe}=2$, but use $\texttt{probe}=4$ on the largest datasets, namely, MS MARCO and Wikipedia. By default we set $\texttt{candidates} = \texttt{probe} * 2^{12}$, but for Wikipedia we set $\texttt{candidates} = \texttt{probe} * 2^{13}$ and for MS MARCO we set $\texttt{candidates} = \texttt{probe} * 2^{14}$. We leave extensive tuning of hyperparameters to future work.

We train on MS MARCO using 64-way tuples for distillation, sampling them from the top-500 retrieved passages per query. The training set of MS MARCO contains approximately 800k queries, though only about 500k have associated labels. We apply distillation using all 800k queries, where each training example contains exactly one ``positive'', defined as a passage labeled as positive or the top-ranked passage by the cross-encoder teacher, irrespective of its label.

We train for 400k steps, initializing from a pre-finetuned checkpoint using 32-way training examples and 150k steps. To generate the top-$k$ passages per training query, we apply two rounds, following \citet{khattab2021relevance}. We start from a model trained with hard triples (akin to \citet{khattab2021relevance}), train with distillation, and then use the distilled model to retrieve for the second round of training. Preliminary experiments indicate that quality has low sensitivity to this initialization and two-round training, suggesting that both of them could be avoided to reduce the cost of training. %

Unless otherwise stated, the results shown represent a single run. The latency results in~\secref{fig:latency} are averages of three runs. To evaluate for Open-QA retrieval, we use evaluation scripts from \citet{khattab2021relevance}, which checks if the short answer string appears in the (titled) Wikipedia passage. This adapts the DPR~\cite{karpukhin2020dense} evaluation code.\footnote{\url{https://github.com/facebookresearch/DPR/blob/main/dpr/data/qa_validation.py}} We use the preprocessed Wikipedia Dec 2018 dump released by \citet{karpukhin2020dense}.

For out-of-domain evaluation, we elected to follow \citet{thakur2021beir} and set the maximum document length of ColBERT, RocketQAv2, and \system{} to 300 tokens on BEIR and \dataset{}. \citet{formal2021spladev2} selected maximum sequence length 256 for SPLADEv2 both on MS MARCO and on BEIR for both queries and documents, and we retained this default when testing their system on \dataset{}. Unless otherwise stated, we keep the default query maximum sequence length for \system{} and RocketQAv2, which is 32 tokens. For the ArguAna test in BEIR, as the queries are themselves long documents, we set the maximum query length used by \system{} and RocketQAv2 to 300. For Climate-FEVER, as the queries are relatively long sentence claims, we set the maximum query length used by \system{} to 64.%

We use the open source BEIR implementation\footnote{\url{https://github.com/UKPLab/beir}} and SPLADEv2 evaluation\footnote{\url{https://github.com/naver/splade}} code as the basis for our evaluations of SPLADEv2 and ANCE as well as for BM25 on LoTTE. We use the Anserini~\cite{yang2018anserini} toolkit for BM25 on the Wikipedia Open-QA retrieval tests as in \citet{khattab2021relevance}. We use the implementation developed by the RocketQAv2 authors for evaluating RocketQAv2.\footnote{\url{https://github.com/PaddlePaddle/RocketQA}}

\begin{table*}[t]
\centering
\resizebox{\linewidth}{!}{
\begin{tabular}{@{}lllll@{}}
\toprule
\multicolumn{1}{c}{Topic}   & \multicolumn{1}{c}{Communities} & \multicolumn{1}{c}{\# Passages} & \multicolumn{1}{c}{\# Search queries} & \multicolumn{1}{c}{\# Forum queries} \\ \midrule
\multirow{5}{*}{Writing}    & ell.stackexchange.com           & 108143      & 433               & 1196             \\
                            & literature.stackexchange.com    & 4778        & 7                 & 58               \\
                            & writing.stackexchange.com       & 29330       & 23                & 163              \\
                            & linguistics.stackexchange.com   & 12302       & 22                & 116              \\
                            & worldbuilding.stackexchange.com & 122519      & 12                & 470              \\ \midrule
\multirow{4}{*}{Recreation} & rpg.stackexchange.com           & 89066       & 91                & 621              \\
                            & boardgames.stackexchange.com    & 20340       & 67                & 179              \\
                            & scifi.stackexchange.com         & 102561      & 343               & 852              \\
                            & photo.stackexchange.com         & 51058       & 62                & 350              \\ \midrule
\multirow{8}{*}{Science}    & chemistry.stackexchange.com     & 39435       & 245               & 267              \\
                            & stats.stackexchange.com         & 144084      & 137               & 949              \\
                            & academia.stackexchange.com      & 76450       & 66                & 302              \\
                            & astronomy.stackexchange.com     & 14580       & 15                & 88               \\
                            & earthscience.stackexchange.com  & 6734        & 10                & 50               \\
                            & engineering.stackexchange.com   & 12064       & 16                & 77               \\
                            & datascience.stackexchange.com   & 23234       & 15                & 156              \\
                            & philosophy.stackexchange.com    & 27061       & 34                & 124              \\ \midrule
\multirow{5}{*}{Technology} & superuser.com                   & 418266      & 441               & 648              \\
                            & electronics.stackexchange.com   & 205891      & 118               & 314              \\
                            & askubuntu.com                   & 296291      & 132               & 480              \\
                            & serverfault.com                 & 323943      & 148               & 506              \\
                            & webapps.stackexchange.com       & 31831       & 77                & 55               \\ \midrule
\multirow{11}{*}{Lifestyle} & pets.stackexchange.com          & 10070       & 20                & 87               \\
                            & lifehacks.stackexchange.com     & 7893        & 2                 & 50               \\
                            & gardening.stackexchange.com     & 20601       & 16                & 182              \\
                            & parenting.stackexchange.com     & 18357       & 10                & 87               \\
                            & crafts.stackexchange.com        & 3094        & 4                 & 50               \\
                            & outdoors.stackexchange.com      & 13324       & 16                & 76               \\
                            & coffee.stackexchange.com        & 2249        & 11                & 50               \\
                            & music.stackexchange.com         & 47399       & 65                & 287              \\
                            & diy.stackexchange.com           & 82659       & 135               & 732              \\
                            & bicycles.stackexchange.com      & 35567       & 40                & 229              \\
                            & mechanics.stackexchange.com     & 27680       & 98                & 246              \\ \bottomrule
\end{tabular}
}
\caption{Per-community distribution of \dataset{} dev dataset passages and questions.} 
\label{table:lotte_communities}
\end{table*}

\end{document}